\documentclass[aps,prm,twocolumn,superscriptaddress,amsmath,amssymb]{revtex4-2}
\usepackage{graphicx}
\usepackage{dcolumn}
\usepackage{bm}
\usepackage{ marvosym }
\usepackage[utf8]{inputenc}
\usepackage[T1]{fontenc}
\usepackage{mathptmx}
\usepackage{comment}
\usepackage{gensymb}
\newcommand{\RomanNumeralCaps}[1]
    {\MakeUppercase{\romannumeral #1}}
\begin{document}


\title{Stress-tailoring magnetic anisotropy of V$_2$O$_3$/Ni bilayers}


\author{Christian T. Wolowiec}
\email[Corresponding author email: ]{cwolowiec@physics.ucsd.edu}
\affiliation{Department of Physics, University of California, San Diego, La Jolla, California 92093, USA}
\author{Juan Gabriel Ram\'{i}rez}
\affiliation{Department of Physics, Universidad de los Andes, Bogot\'{a} 111711, Colombia}
\author{Min-Han Lee}
\affiliation{Department of Physics, University of California, San Diego, La Jolla, California 92093, USA}
\affiliation{Materials Science and Engineering Program, University of California, San Diego, La Jolla, California 92093, USA}
\altaffiliation[Present address: ]{Applied Materials, Santa Clara, California 95054, USA}
\author{\mbox{Nicolas M. Vargas}}
\altaffiliation[Present address:  ]{General Atomics, San Diego, California 92121, USA}
\affiliation{Department of Physics, University of California, San Diego, La Jolla, California 92093, USA}
\author{Ali C. Basaran}
\affiliation{Department of Physics, University of California, San Diego, La Jolla, California 92093, USA}
\author{Pavel Salev}
\affiliation{Department of Physics, University of California, San Diego, La Jolla, California 92093, USA}
\author{Ivan K. Schuller}
\affiliation{Department of Physics, University of California, San Diego, La Jolla, California 92093, USA}
\date{\today}

\begin{abstract}
We report on a temperature-driven reversible change of the in-plane magnetic anisotropy of  V$_2$O$_3$/Ni bilayers. This is caused by the rhombohedral to monoclinic structural phase transition of V$_2$O$_3$ at $T_C$ = 160 K. The in-plane magnetic anisotropy is uniaxial above $T_C$, but as the bilayer is cooled through the structural phase transition, a secondary magnetic easy axis emerges. Ferromagnetic resonance measurements show that this change in magnetic  anisotropy is reversible with temperature. We identify two structural properties of the V$_2$O$_3$/Ni bilayers affecting the in-plane magnetic anisotropy: (1) a growth-induced uniaxial magnetic anisotropy associated with step-like terraces in the bilayer microstructure and (2) a low-temperature strain-induced biaxial anisotropy associated with the V$_2$O$_3$ structural phase transition. Magnetoresistance measurements corroborate the change in magnetic anisotropy across the structural transition and suggest that the negative magnetostriction of Ni leads to the emergence of a strain-induced easy-axis. This shows that a temperature-dependent structural transition in V$_2$O$_3$ may be used to tune the magnetic anisotropy in an adjacent ferromagnetic thin film.

\end{abstract}


\maketitle
\section{Introduction}
\label{Introduction}
The magnetic properties of most materials are strongly dependent on lattice strain. This is especially important for the growth and use of multilayer heterostructures in magnetic devices and applications in which the individual layers can be subject to appreciable stress, owing to a number of causes, including lattice mismatch between adjacent layers. Hence, a successful device that is based on a strained heterostructure relies on the ability to control both the morphology and state of strain in an individual layer. A well-known method for inducing an in-plane uniaxial magnetic anisotropy in a ferromagnetic film is to grow the film on a surface with a terraced microstructure, where an in-plane magnetic easy axis (EA) typically emerges in a direction parallel or perpendicular to the terrace boundaries \cite{Albrecht_1992, Metoki_1993, Chuang_1994, Kawakami_1996, Liu_2013, Davydenko_2014, Metoki_1993}. Strain is another route for modifying the in-plane magnetic anisotropy in a thin ferromagnetic film \cite{Kittel_1949, Lee_1955, Liu_2014}. However, surface morphology and the state of strain are typically introduced during growth with little or no opportunity for post-synthesis modulation of the magnetic properties of the ferromagnetic film. Recent investigations of the epitaxial growth of ferromagnetic Ni thin films on vanadium oxides demonstrate post-synthesis changes (in addition to growth-induced changes) to the magnetic structure in the Ni layer \cite{Gilbert_2017, delaVenta_2013, delaVenta_2014}. 

Vanadium sesquioxide (V$_2$O$_3$) is well known for its triple-coincident electronic, structural, and magnetic transitions where the high-temperature (HT) rhombohedral, metallic, and paramagnetic (PM) phases undergo transitions to the low-temperature (LT) monoclinic, insulating, and antiferromagnetic (AFM) phases upon cooling through a transition temperature of $T_C$ = 160 K \cite{Kalcheim_2019, Kosuge_1967, Dernier_1970, Ueda_1980, Singer_2018}. As a result of the structural change, the volume of the V$_2$O$_3$ unit cell expands approximately 1.4\% upon cooling into the monoclinic phase, during which there is an increase in the $a$ lattice parameter from  from 2.87 to 2.91 \AA~\cite{Singer_2018}. This expansion can produce an in-plane tensile strain in a thin-film that is grown adjacent to the V$_2$O$_3$ layer. Numerous reports show that the structural changes in V$_2$O$_3$ at $T_C$ significantly affect the coercive fields in neighboring ferromagnetic thin films \cite{Gilbert_2017, delaVenta_2013, delaVenta_2014, Ramirez_2016, Saerbeck_2014}. Moreover, direct observation of the change in the Ni spin alignment across the SPT in V$_2$O$_3$ was recently reported in a V$_2$O$_3$/Ni bilayer \cite{Valmianski_2021}. However, only a general description of a proximity-induced strain has been given as the cause for the magnetic changes in the neighboring ferromagnetic Ni layer. Specifically, there is still no clear understanding of the mechanism responsible for the emergence of a new anisotropy axis in V$_2$O$_3$/Ni bilayers across the first-order phase transition in V$_2$O$_3$.  

Here, we track the temperature-dependent changes to the in-plane magnetic anisotropy and magneto-transport in V$_2$O$_3$/Ni bilayer films across the V$_2$O$_3$ phase transition. By measuring the ferromagnetic resonance (FMR) and the anisotropic magnetoresistance (AMR) of two V$_2$O$_3$/Ni bilayer samples, we were able to distinguish between two sources of crystallographic strain: (1) a relatively weak strain in the Ni layer associated with a stressed V$_2$O$_3$  layer that is caused by a misalignment of the V$_2$O$_3$ (012) plane with the sapphire Al$_2$O$_3$ (012) substrate during growth and (2) a strong strain in the Ni layer related to the rhombohedral to monoclinic structural phase transition in V$_2$O$_3$ upon cooling through $T_C$. We present a scenario in which the strain in the ferromagnetic Ni layer is coupled to the terraced microstructure of the V$_2$O$_3$ layer grown on sapphire. Here, we propose that elongated ``rips'' in the  bilayer, caused by step-like terraces in the substrate \cite{Gilbert_2017}, ``funnel'' the strain into the regions of Ni that are parallel to the terraced boundaries. Hence, the in-plane strain in the Ni layer is effectively a uniaxial strain directed along the terrace boundaries. Due to the negative magnetostriction coefficient of Ni, this in-plane tensile strain that is directed along the terrace boundaries causes a 90\degree~reversal of magnetic domains within the Ni layer. This scenario is consistent with the emergence of a secondary anisotropy axis across the SPT in V$_2$O$_3$ as observed in the FMR, in which there is a change from a HT uniaxial magnetic anisotropy (with one easy axis parallel to the terrace boundaries) to a LT biaxial magnetic anisotropy with the emergence of a secondary easy axis that is perpendicular to the terrace boundaries. 

As a measure of the spin-dependent scattering, the in-plane AMR corroborates the FMR and the re-orientation of the magnetic domains in the Ni layer. In sufficiently strong field, AMR measurements confirm the typical $\cos^2$ dependence observed for the electrical resistance of polycrystalline ferromagnetic thin films \cite{McGuire_1975}. A comparison of the spin-dependent scattering in strong field with that at the coercive field $H_c$ suggests that the spontaneous orientation of the magnetic domains in the Ni layer align parallel to the terrace boundaries. Analysis of the AMR at temperatures above and below the structural change in V$_2$O$_3$ indicates that there is a 90\degree~strain-induced reversal of magnetic domains within the Ni layer that is consistent with the emergence of the secondary easy axis observed in the low-temperature FMR.

\section{Experimental}
\label{Experimental}
\begin{figure}[t]
   \includegraphics[scale = 0.7, trim= 7.5cm 4cm 0cm 4cm, clip=true]{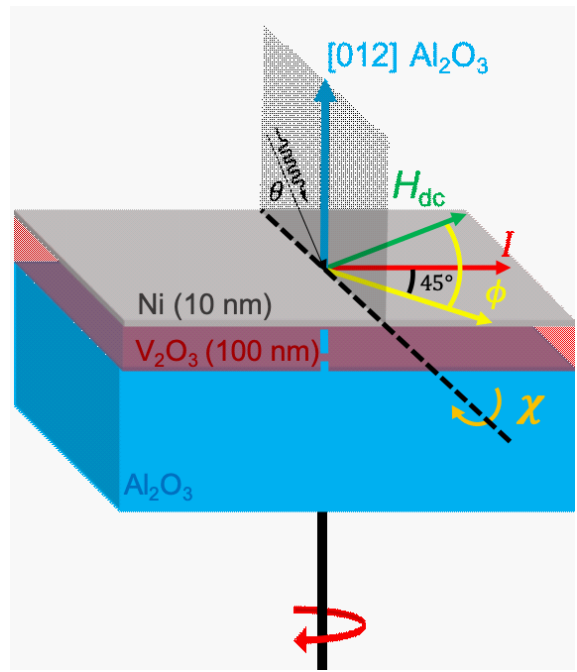}
\caption{\label{fig1} (Color online) Schematic representation of the V$_2$O$_3$/Ni bilayer on an \textit{r}-cut sapphire (Al$_2$O$_3$) substrate showing (1) the geometry for the XRD  and RSM crystallographic measurements and (2) the in-plane geometry of the FMR and AMR measurements. The X-ray beam is aligned to the Al$_2$O$_3$ (012) plane. The angle $\theta$ represents the angle between the beam and the Al$_2$O$_3$ (012) plane and the angle $\chi$ represents the tilting angle. For the in-plane FMR measurements, the angle between the in-plane dc magnetic field $H_{dc}$ (green arrow) and the magnetic easy axis direction (yellow arrow) is given by $\phi$. For the in-plane AMR measurements, the initial angle of 0\degree~between $H_{dc}$ and the electric current $I$ (red arrow) corresponds to $\phi = 45$\degree. (The angle between the electric current $I$ and the easy axis direction is fixed at 45\degree.)} 
\end{figure}

Two thin-film V$_2$O$_3$/Ni bilayers, V$_2$O$_3$(100 nm)/Ni(10 nm)/Pt(3 nm) and V$_2$O$_3$(100 nm)/Ni(10 nm)/Al(3 nm), were separately grown on \textit{r}-cut (012) sapphire (Al$_2$O$_3$) substrates in a high-vacuum sputtering chamber with a base pressure of $\sim$ 2 $\times$ $10^{-7}$ Torr. The 100 nm V$_2$O$_3$ films were deposited by RF magnetron sputtering at 150 W using a homemade V$_2$O$_3$ stoichiometric target. During the deposition of V$_2$O$_3$, the sapphire substrates were held at $\sim$ 700 \degree C in a 7.8 mTorr Ar atmosphere. After quenching the V$_2$O$_3$ layers to room temperature and recovering the base pressure, the 10 nm Ni films were deposited onto the V$_2$O$_3$ layers by RF magnetron sputtering at 100 W using an elemental Ni target at room temperature in a 4 mTorr Ar atmosphere. The structure of both samples, obtained from room temperature X-ray diffraction (XRD) and reciprocal space maps (RSMs) confirm epitaxial single-phase growth of V$_2$O$_3$. The geometry of the X-ray measurements is shown in the schematic displayed in Fig.~\ref{fig1}. 
In-plane ferromagnetic resonance (FMR) measurements of the V$_2$O$_3$/Ni bilayers were performed upon cooling from 296 to 100 K using a Bruker BioSpin electron paramagnetic resonance spectrometer with a cylindrical cavity resonator at a fixed frequency \textit{f} = 9.4 GHz and a variable dc magnetic field \textit{H}. The sample was mounted on a quartz rod and rotated through the angle $\phi$ at two-degree intervals within an accuracy of $\Delta$$\phi=$ 0.025\degree. At each angle $\phi =$ 0\degree,2\degree,4\degree,~\ldots,360\degree, the applied, in-plane dc magnetic field was swept from 9000 to 0 Oe at a rate of 150 Oe/s, where the in-plane angle $\phi$ is defined as the angle between the in-plane dc magnetic field H and the uniaxial magnetic easy ($\phi=$ 0\degree). To increase the signal to noise ratio, all FMR measurements were performed with the microwave power kept constant at 1 mW, which avoids any appreciable sample heating. (Results of FMR measurements at higher power are shown in Fig.~\ref{figS3} in the Supplemental Material.)  The resonance field $H_R$ was determined from a fit of a dynamical model (described in the Supplemental Material) to the FMR signal at each angle $\phi =$ 0\degree, 2\degree, 4\degree,~\ldots, 360\degree.  
\begin{figure*}[t]
   \includegraphics[scale = 0.41, trim= 0cm 0.2cm 13cm 0cm, clip=true]{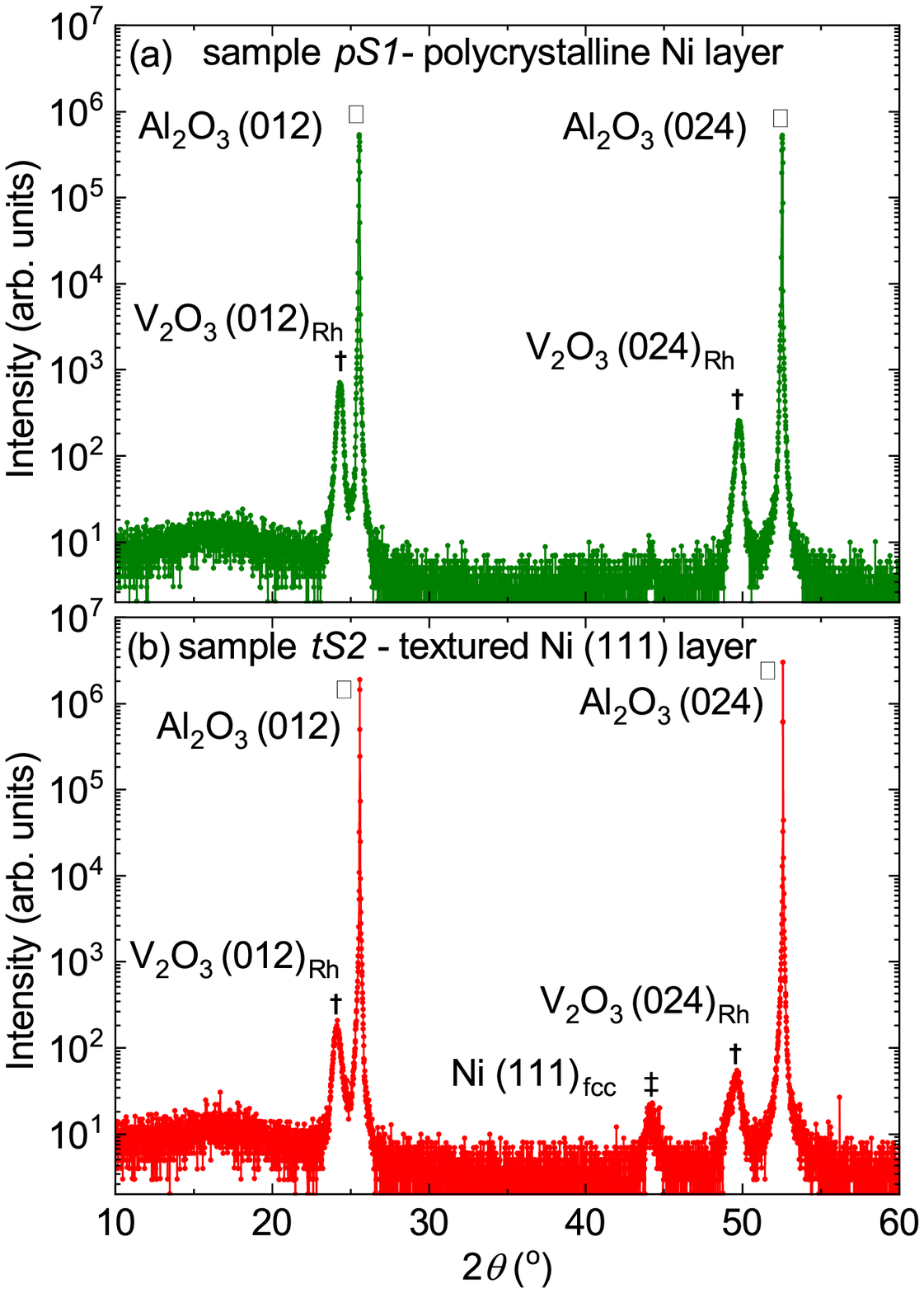}
   \hspace{-0.4cm}
    \includegraphics[scale = 0.42, trim= 0.5cm 0.4cm 12cm 0cm, clip=true]{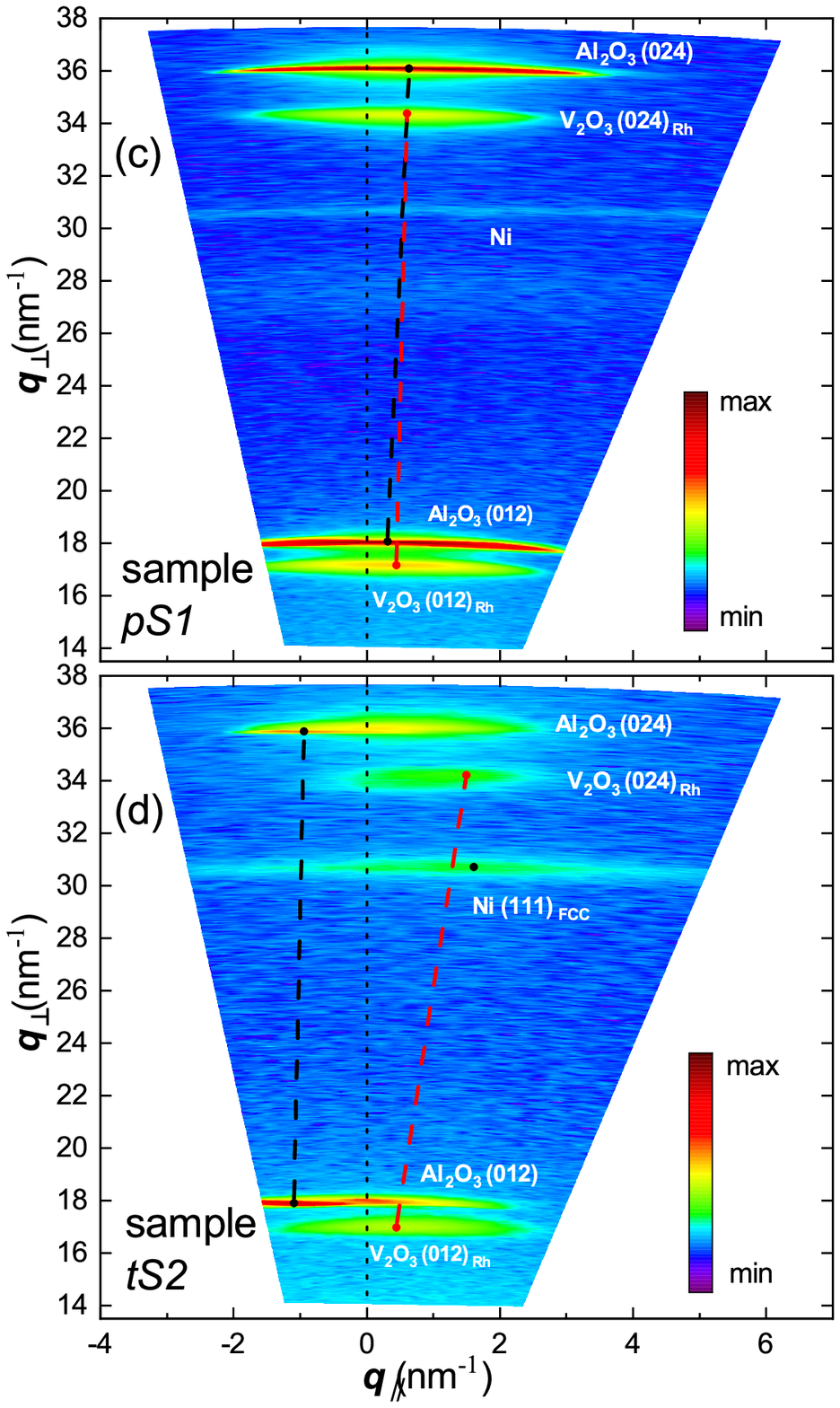}
    \hspace{-0.9cm}
     \includegraphics[scale = 0.4, trim= 8cm 2cm 6cm 4cm, clip=true]{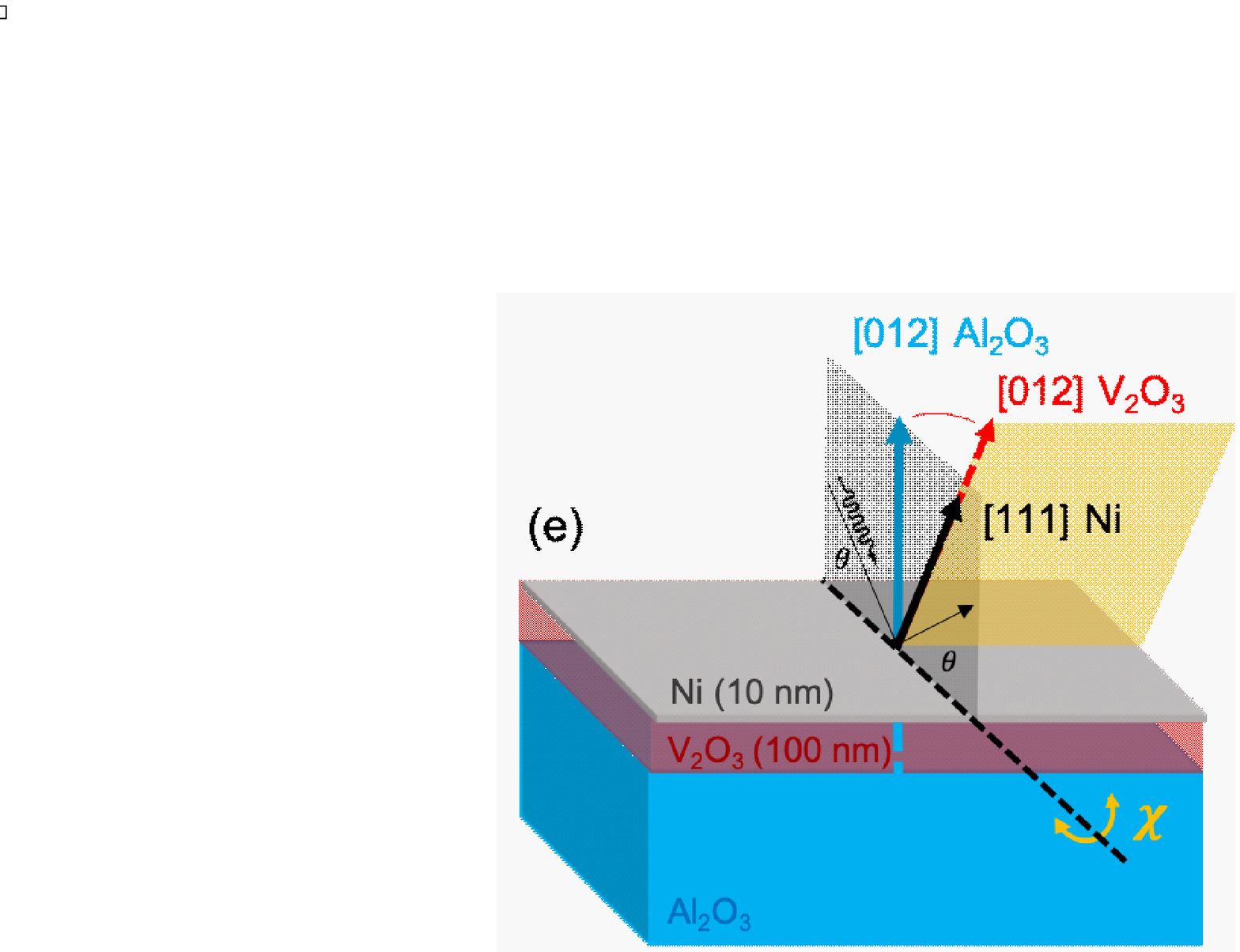}
\caption{\label{fig2} (Color online) (a, b) Room-temperature X-ray diffraction (XRD) for the two different V$_2$O$_3$(100 nm)/Ni(10 nm) bilayer samples (a) $pS1$ and (b) $tS2$. The Bragg peaks from the Al$_2$O$_3$ (012) and (024) planes are marked by $\ast$, the V$_2$O$_3$ (012)$_{Rh}$ and (024)$_{Rh}$ peaks are marked by \Cross~and the Ni (111) peak in panel (b) is marked by \textbf{\ddag}. (c, d) Reciprocal space maps (RSMs) for the two V$_2$O$_3$/Ni bilayer samples (c) $pS1$ and (d) $tS2$ with the X-ray beam aligned to the Al$_2$O$_3$ (012) plane. (e) Schematic representation of the V$_2$O$_3$/Ni bilayer on an \textit{r}-cut sapphire (Al$_2$O$_3$) substrate showing the relative orientation of the Ni (111), V$_2$O$_3$ (012)$_{Rh}$ and Al$_2$O$_3$ (012) planes. The angle $\theta$ represents the angle between the beam and the Al$_2$O$_3$ (012) plane and the angle $\chi$ represents the tilting angle.} 
\end{figure*}
\indent In-plane anisotropic magnetoresistance (AMR) measurements of one V$_2$O$_3$/Ni bilayer sample (with a textured Ni(111) layer) were performed upon warming at temperatures 100, 150, and 300 K. Annealed Pt wire leads were affixed to the surface of the V$_2$O$_3$/Ni bilayer sample with silver epoxy in a standard four-wire configuration so that the electric current is directed at a fixed angle of 45\degree~relative to the uniaxial magnetic easy axis as shown in the schematic displayed in Fig.1. The electrical resistance R was measured using a Quantum Design physical property measurement system with the dc magnetic field \textit{H} applied parallel to the surface of the V$_2$O$_3$/Ni bilayer. The angle between the electric current and the direction of the applied magnetic field was varied in ten degree intervals. At each angle, the electrical resistance \textit{R} was measured as the magnitude of  \textit{H} was swept from +1000 Oe to $-$1000 Oe and then back to +1000 Oe. Due to the orientation of the electric current $I$ relative to the uniaxial magnetic easy axis, the initial angle of  0\degree~in the magnetoresistance measurement corresponds to $\phi = $ 45\degree~in the FMR measurement (see Fig.~\ref{fig1}).
\section{Results}
\label{Results}
X-ray measurements were performed on two different V$_2$O$_3$/Ni bilayer samples grown on \textit{r}-cut sapphire (Al$_2$O$_3$) substrates. X-ray diffraction (XRD) and reciprocal space maps (RSMs) shown in Fig.~\ref{fig2} confirm the epitaxial growth of a single-phase (rhombohedral) V$_2$O$_3$ layer in both samples. The peak at 2$\theta$ $\approx$ 44.2\degree~in the XRD pattern of Fig.~\ref{fig2}(b) is indicative of a textured face-centered-cubic (fcc) Ni layer with the Ni [111] direction shown in Fig.~\ref{fig2}(e) to be nearly parallel to the V$_2$O$_3$ [012] direction. The maximum in the intensity in the RSM shown in Fig.~\ref{fig2}(d) (indicated by the black dot located at ($\vec{q}_{\parallel}$, $\vec{q}_{\perp}$) $\approx$ (1.5 nm$^{-1}$, 30 nm$^{-1}$)) is further evidence of a textured Ni (111) layer in this sample. In contrast, the absence of a Ni peak in the XRD pattern of Fig.~\ref{fig2}(a) and the absence of a localized intensity maximum in the RSM shown in Fig.~\ref{fig2}(c) together suggest the Ni layer is polycrystalline in this sample.\\ 
\indent A comparison of the two RSMs in Fig.~\ref{fig2} shows that while the family of V$_2$O$_3$ (012) planes is nearly parallel to the Al$_2$O$_3$ substrate in one sample (Fig.~\ref{fig2} (c)), the V$_2$O$_3$ (012) planes are oblique to the Al$_2$O$_3$ substrate in the other sample (Fig.~\ref{fig2}(d)). This is illustrated by the dashed black and red lines drawn in the RSMs of Figs.~\ref{fig2}(c), (d). Heavy dashed black lines connect the intensity maxima for the Al$_2$O$_3$ (012) and (024) planes, while the heavy dashed red lines connect intensity maxima for the V$_2$O$_3$ (012) and (024) planes. The nearly overlapping dashed lines in Fig.~\ref{fig2}(c) indicate the family of V$_2$O$_3$ (012) and (024) planes are nearly parallel to the substrate. In contrast, the separated dashed lines in Fig.~\ref{fig2}(d) indicate the V$_2$O$_3$ planes are significantly tilted with respect to the Al$_2$O$_3$ (012) plane of the substrate. A schematic representation of the V$_2$O$_3$/Ni bilayer on \textit{r}-cut sapphire (Al$_2$O$_3$) substrate shown in Fig.~\ref{fig2}(e) illustrates the relative orientation of the V$_2$O$_3$ (012) and Al$_2$O$_3$ (012) planes. Figure~\ref{fig2}(e) can be compared with the geometry of the RSM measurements shown in Fig.~\ref{fig1} in the Experimental section. 
\begin{figure*}[t]
   \includegraphics[scale = 0.39, trim= 2cm 0.9cm 4cm 0.5cm, clip=true]{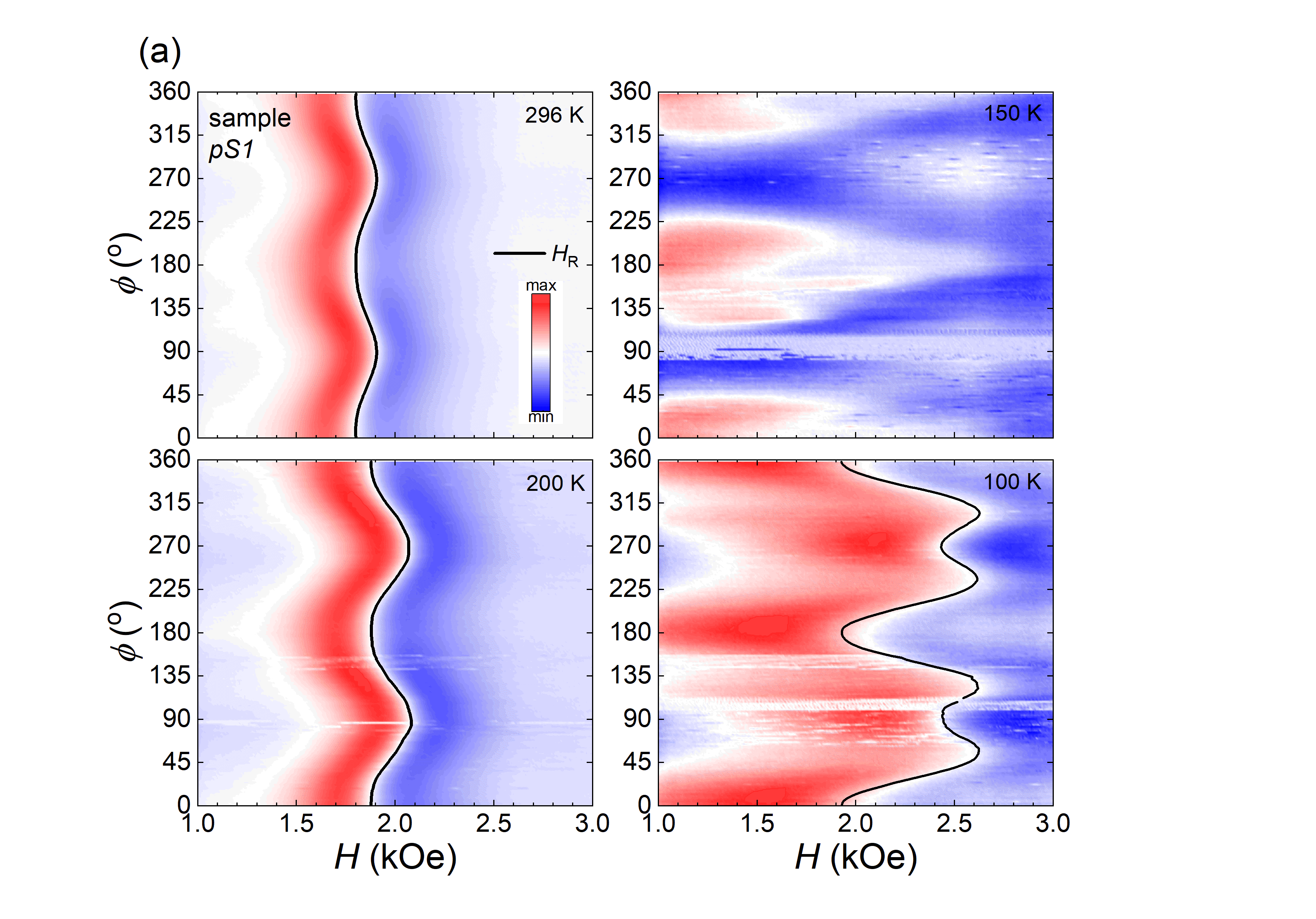}
  \hspace{-0.7 cm}
    \includegraphics[scale = 0.39, trim= 2cm 0.9cm 4cm 0.5cm, clip=true]{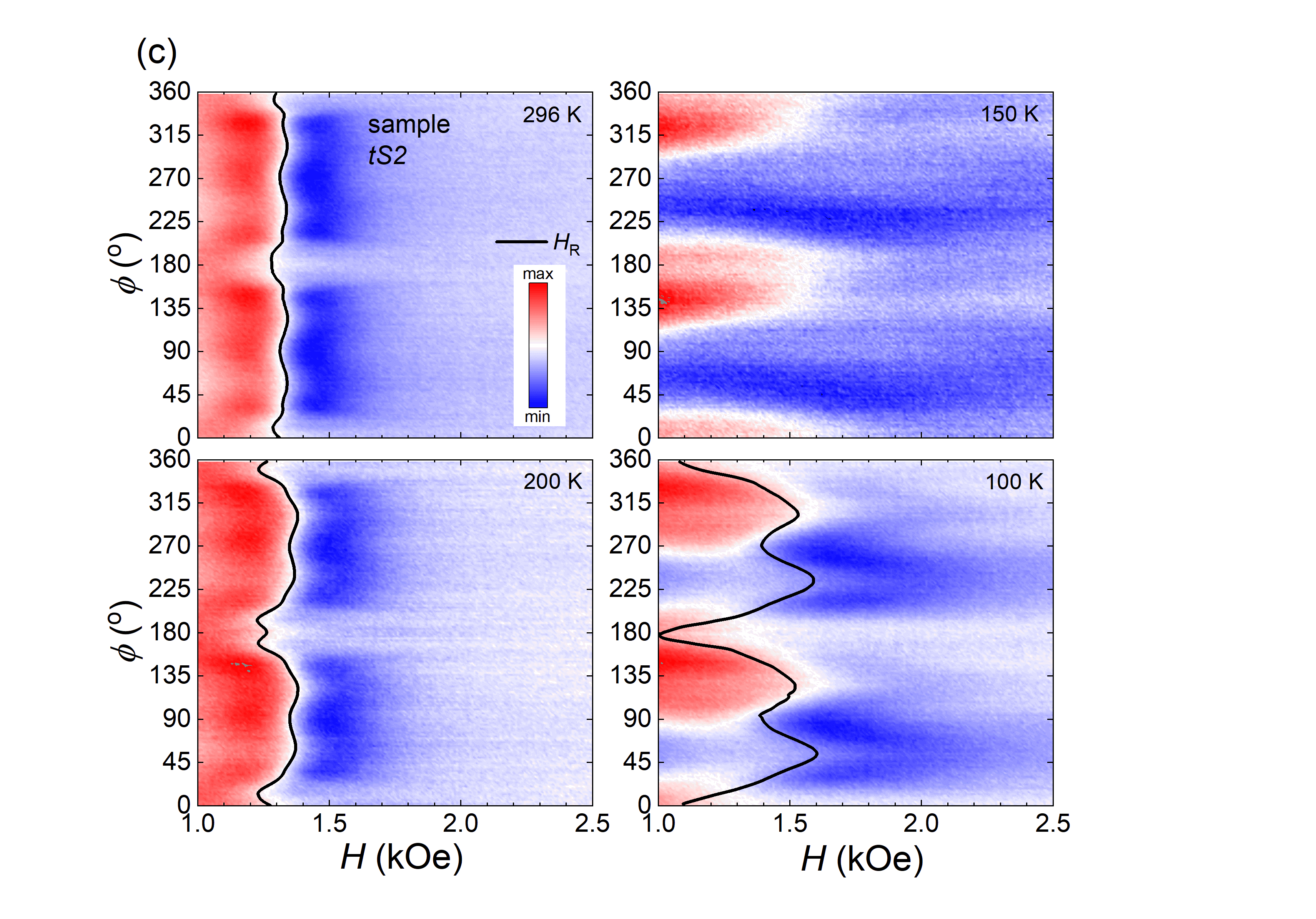}
     \includegraphics[scale = 0.3, trim= 1.3cm 1cm -1cm 1cm, clip=true]{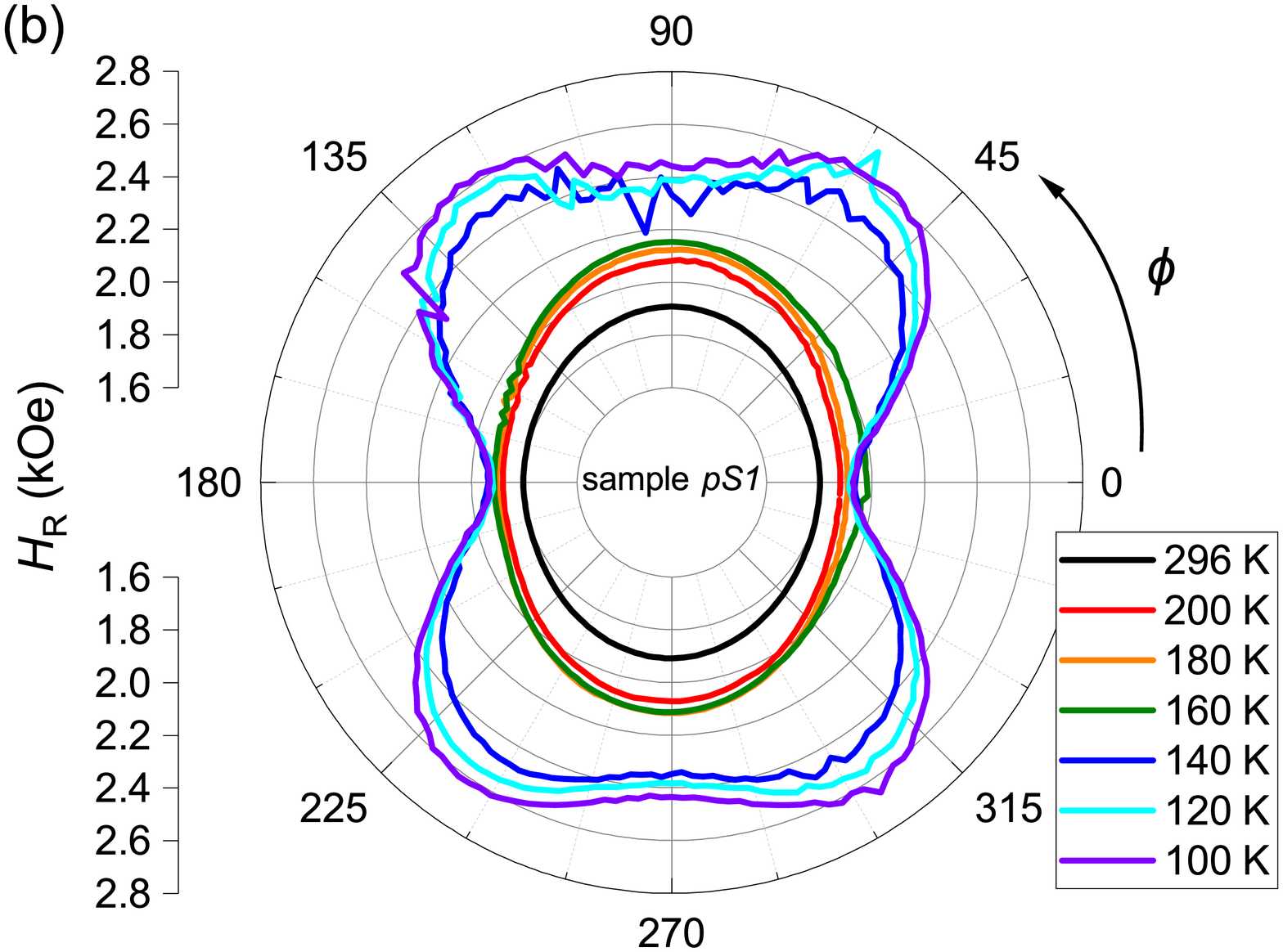}
     \includegraphics[scale = 0.3, trim= 0cm 1cm 2cm 1cm, clip=true]{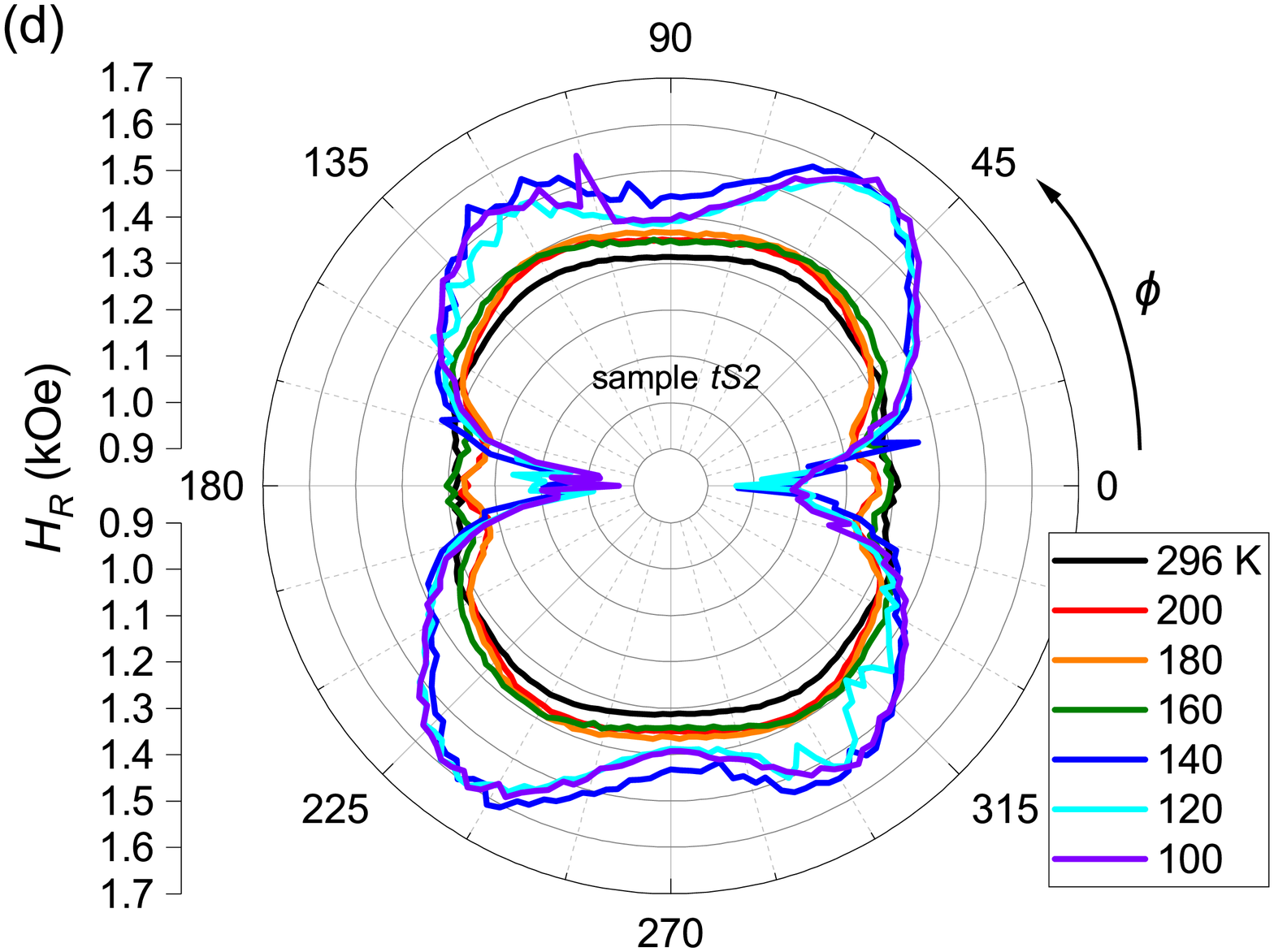}     
\caption{\label{fig3} (Color online) In-plane angular-dependent ferromagnetic resonance (FMR) measurements of two V$_2$O$_3$/Ni bilayer samples. Contour plots of the in-plane FMR signal as a function of dc magnetic field $H_{dc}$ at various temperatures above and below the V$_2$O$_3$ SPT for the (a) \textit{pS1} and (c) \textit{tS2} samples. The y-axis variable $\phi$ is the angle between $H_{dc}$ and the magnetic easy axis. Blue to red contrast indicates low to high signal intensity. The angular dependence of the resonance field $H_{R}$ (solid black curve) was extracted from a fit of a dynamical model to the FMR. The data shown was taken upon cooling from 296 to 100 K. Warming data (not shown) taken at the same temperatures exhibits nearly identical behavior. (b),(d) Polar representations of the angular dependence of $H_{R}$ from 296 to 100 K for the (b) \textit{pS1} and (d) \textit{tS2} samples.} 
\end{figure*}
The real space orientation of the (012) crystallographic planes for V$_2$O$_3$ and Al$_2$O$_3$ are nearly parallel ($\chi$ $\approx$ 0.5\degree) for the sample with polycrystalline Ni (Fig.~\ref{fig2}(c)) but are misaligned at a significant angle of ($\chi$ $\approx$ 5\degree) for the sample with a (111) textured Ni layer (Fig.~\ref{fig2}(d)). For convenience,  \textit{pS1} will refer to the V$_2$O$_3$/Ni bilayer sample with a polycrystalline Ni layer deposited on a ``flat'' V$_2$O$_3$ (012) plane, while  \textit{tS2} will refer to the sample with a textured Ni layer deposited on an ``inclined'' V$_2$O$_3$ (012) plane.\\     
\indent In-plane angular-dependent ferromagnetic resonance (FMR) measurements were performed on two V$_2$O$_3$/Ni bilayer samples as a function of temperature above and below the V$_2$O$_3$ transition at $T_C$ = 160 K. The evolution of the in-plane magnetic anisotropy of sample  \textit{pS1} is shown at select temperatures in Fig.~\ref{fig3}(a). The solid black curves superimposed on the contour plots represent the angular-dependent resonance field $H_R$. The value of $H_R$ was determined from a fit of a dynamical model (described in the Supplemental Material) to the FMR signal at each angle $\phi$ = 0\degree, 2\degree, 4\degree,~\ldots, 360\degree.  The angle $\phi$ is defined as the angle between the applied, in-plane dc magnetic field $H$ and the uniaxial easy axis (see Fig.~\ref{fig1} in the Experimental section). The high-temperature (HT) magnetic anisotropy (at 296 K and 200 K) in the  \textit{pS1} sample (Fig.~\ref{fig3}(a)) is completely uniaxial with one easy axis (EA) at $\phi$ = 0\degree. The HT magnetic anisotropy in sample  \textit{tS2} (Fig.~\ref{fig3}(c)) is a superposition of a uniaxial anisotropy (also with an EA at $\phi$ = 0\degree) and a weak biaxial anisotropy with a secondary EA at $\phi$ = 90\degree. (See Figs.~\ref{figS1} and~\ref{figS2} in the Supplemental Material for a complete set of angular-dependent FMR data for both the  \textit{pS1} and  \textit{tS2} samples at temperatures down to 100 K).\\
\begin{figure*}[t]
   \includegraphics[scale = 0.6, trim= 0.2cm 0.2cm 0.2cm 0cm, clip=true]{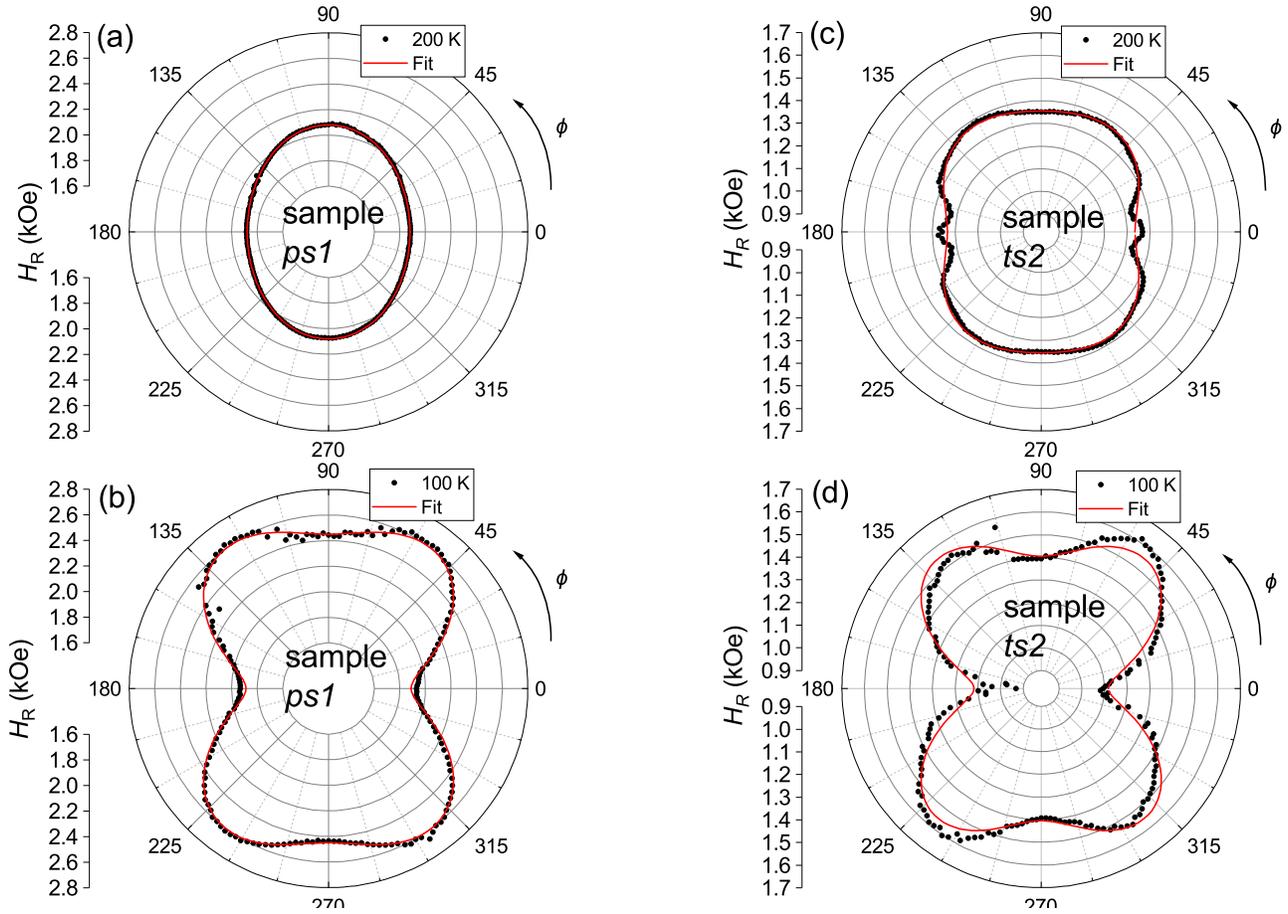}
    \caption{\label{fig4} (Color online) Representative fits (red curves) of the anisotropy energy $E_a$ (see Equation~\ref{equation1} in text) to the resonance field lines $H_R$($\phi$) (black circles) for the \textit{pS1} sample at (a) 200 K and (b) 100 K and for the \textit{tS2} sample at (c) 200 K and (d) 100 K. (See Figs.~S4 and S5 in the Supplemental Material for the fits at different temperatures.)} 
\end{figure*}
\indent Upon cooling, the HT magnetic anisotropy in both samples vanishes near the first-order phase transition in V$_2$O$_3$ at 150 K, where phase coexistence leads to an overall magnetically disordered state of the sample. The magnetic anisotropy reemerges in an altered form at low temperature (LT) below the SPT in V$_2$O$_3$ (see contour plots at 150 K and 100 K in Fig.~\ref{fig3}(a) and Fig.~\ref{fig3}(c)). The magnetic anisotropy in both samples at LT exhibits a superposition of a uniaxial magnetic anisotropy (with one EA at $\phi$ = 0\degree) and a biaxial anisotropy (with one EA at $\phi$ = 0\degree~and a secondary EA at $\phi$ = 90\degree). Both samples exhibit a LT biaxial magnetic anisotropy similar to  the biaxial anisotropy at HT (296 K and 200 K) observed in the  \textit{tS2} sample with a misaligned V$_2$O$_3$ layer. This is shown in the angular dependence of $H_R$ for all FMR measurements down to 100 K, summarized in the polar plots shown in Fig.~\ref{fig3}(b) for  \textit{pS1} and Fig.~\ref{fig3}(d) for \textit{tS2}.  The elliptically shaped plots of $H_R$($\phi$) shown in Fig.~\ref{fig3}(b) indicate a purely uniaxial magnetic anisotropy down to a temperature of 160 K for sample  \textit{pS1}. In contrast, the angular dependence of $H_R$ for sample  \textit{tS2} at HT exhibits a very weak biaxial character superposed on a predominantly uniaxial magnetic anisotropy (Fig.~\ref{fig3}(d)). Below the V$_2$O$_3$ triple transition at $T_C$ = 160 K, there is a clear departure in both samples from the predominant uniaxial anisotropy at HT to a biaxial anisotropy at LT. \\
\indent It is well established that the total magnetic anisotropy energy $E_a$ is determined by the intrinsic magnetocrystalline anisotropy and the extrinsic contributions related to strain (or magnetoelastic anisotropy), shape anisotropy, exchange anisotropy, and the Zeeman energy cite \cite{Cullity_2008, Coey_2010a, Coey_2010b}. In the two V$_2$O$_3$/Ni bilayer samples investigated here, the magneto-elastic anisotropy is the predominant in-plane contribution. For both the  \textit{pS1} and  \textit{tS2} samples, the FMR indicates that the in-plane magnetic anisotropy is either uniaxial or a superposition of uniaxial and biaxial anisotropies. Hence, the in-plane magnetic anisotropy energy $E_a$ for the V$_2$O$_3$/Ni bilayers in this report can be represented by the following expression \cite{Urban_2001}: 
\begin{equation} 
\label{equation1}
                            E_a = K_0 + K_Ucos^2(\phi-\phi_1) + K_{ME}sin^2(2(\phi-\phi_2)).             
\end{equation}
The zero point of the magnetic anisotropy energy is determined by the constant $K_0$. The second term represents the energy associated with the uniaxial magnetic anisotropy, with a hard axis at an angle $\phi_1$ = 90\degree~relative to the EA. The third term represents the energy associated with the biaxial magnetic anisotropy with a hard axis at an angle $\phi_2$ = 45\degree~relative to the easy axis \cite{Urban_2001}. The anisotropy constants $K_U$ and $K_{ME}$  were determined from fits of the resonance field $H_R$($\phi$) to the expression for the magnetic anisotropy energy $E_a$ given in Equation (1). The determination of $K_U$ and $K_{ME}$ required a conversion of $H_R$ to a magnetic energy density $u_m$ = $BH_R/2$ in units of J/m$^3$, where $B$ = $\mu_0$ ($H_R$ + $M_S$) and $M_S$ = 485 emu/cm$^3$ is taken as the saturation magnetization of Ni \cite{Tannous_2008}. Selected fits to $H_R$ at 200 K and 100 K for the  \textit{pS1} and  \textit{tS2} samples are shown in Figs.~\ref{fig4}(a), (b), (c), and (d). (See Fig.~\ref{figS4} and Fig.~\ref{figS5} in the Supplemental Material for the complete set of fits at other temperatures.)\\
\indent The temperature dependence of the anisotropy constants $K_U$ and $K_{ME}$ are plotted in Fig.~\ref{fig5}(a).  Both anisotropy constants are sensitive to temperature and exhibit significant changes in magnitude upon cooling through the SPT in V$_2$O$_3$ at 160 K. In the Ni layer of the  \textit{pS1} sample with an ``unstressed'' V$_2$O$_3$ layer at HT, the magnetic anisotropy is purely uniaxial above the SPT, as indicated by the constant value of $K_{ME}$ near 0 J/m$^3$ (black squares). (The small positive value for $K_{ME}$ that resulted from the fit is unphysical and reflects the absence of strain in the  \textit{pS1} sample.)  The sign of $K_{ME}$ for this sample changes from positive to negative at the onset of the induced strain across the SPT in V$_2$O$_3$ at 160 K. In contrast, $K_{ME}$  is initially negative at HT in the \textit{tS2} sample (open squares) and exhibits a smaller temperature dependence upon cooling to the SPT at 160 K. The small but non-zero magnitude of $K_{ME}$ at HT in this V$_2$O$_3$/Ni bilayer reflects the presence of a slight strain in the Ni layer. Moreover, the slight temperature dependence of $K_{ME}$ down to the SPT as displayed in Fig.~\ref{fig5}(a) is also representative of the weak biaxial anisotropy at HT observed in the FMR for this  \textit{tS2} sample (see Fig.~\ref{fig3}(c), (d)).  
\begin{figure}[t]
   \includegraphics[scale = 0.33, trim= 1.2cm 3.2cm 0cm 2cm, clip=true]{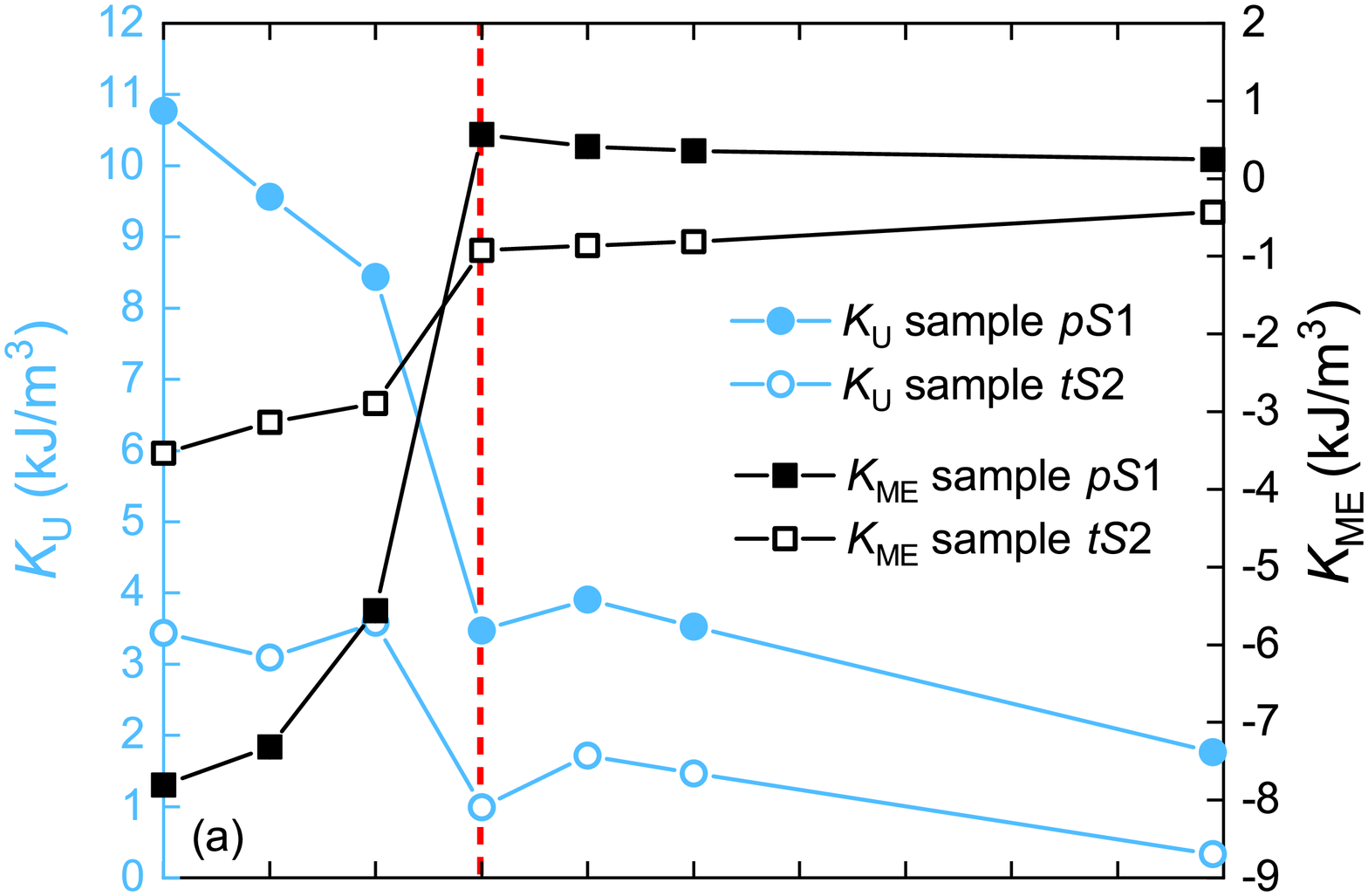}
    \includegraphics[scale = 0.33, trim= 1.2cm 1.3cm 0cm 2cm, clip=true]{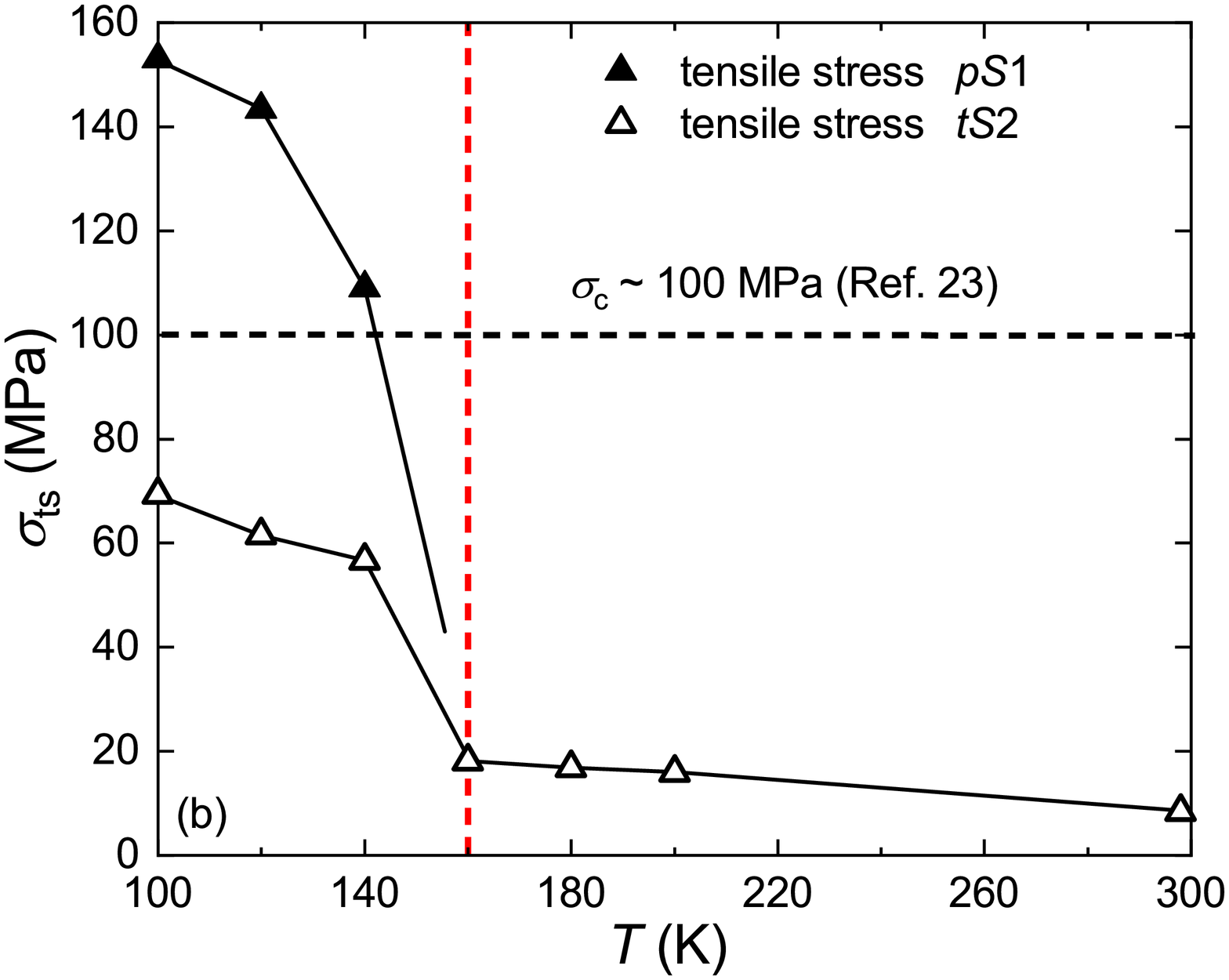}
\caption{\label{fig5} (Color online) (a) The temperature dependence of the magnetic anisotropy constants $K_U$ and $K_{ME}$ for the two V$_2$O$_3$/Ni bilayer samples \textit{pS1} (filled symbols) and \textit{tS2} (open symbols). The anisotropy constants $K_U$ and $K_{ME}$ were extracted from fits of Equation~\ref{equation1} to the $H_R$ curves determined from the FMR measurements. The dashed red line marks the V$_2$O$_3$ transition temperature $T_C$. (b) Tensile stress vs.~temperature ($\sigma_{ts}$ vs.~$T$) for samples \textit{pS1} (solid triangles) and \textit{tS2} (open triangles). The horizontal dashed line indicates the critical stress $\sigma_c$ = 100 MPa of ferromagnetic Ni \cite{Cullity_2008}. The extrapolation of $\sigma_{ts}$ to 0 MPa for sample $pS1$ is based on the value of $K_{ME}$ $\approx$ 0 kJ/$m^3$ in panel (a).} 
\end{figure}
The magnitude of $K_{ME}$ jumps significantly for both samples at the onset of the SPT (vertical dashed red line in Fig.~\ref{fig5}(a)), indicating a significant proximity-induced strain in the Ni layer at the SPT in V$_2$O$_3$. The jump in $K_U$ across the SPT is somewhat weaker for both samples. For the Ni layer of the  \textit{pS1} sample (light-blue circles), $K_U$ increases by a factor of 2 to 3 across the SPT from 160 to 140 K. For the Ni layer of the \textit{tS2} sample (open-blue circles), there is hardly any deviation in the extrapolation of the temperature dependence of $K_U$ compared to the jump in $K_U$ across the SPT. This weaker relative change in the uniaxial anisotropy $K_U$ across the SPT is consistent with the expectation that the microstructure does not change significantly due to strain coming from the SPT in the underlying V$_2$O$_3$ layer. The magnitude of $K_{ME}$  can be used to calculate the tensile stress $\sigma_{ts} = \frac{2K_{ME}}{3\lambda_{si}}$ \cite{Cullity_2008b}, where $\lambda_{s}$ $\approx$ $-$34$\times$10$^{-6}$ is the negative magnetostriction coefficient for a polycrystalline Ni layer \cite{Cullity_2008b}. Figure~\ref{fig5}(b) displays the tensile stress $\sigma_{ts}$ in the Ni layer as a function of temperature. As expected, there is a significant increase in tensile stress $\sigma_{ts}$ below the SPT at $T_C$ = 160 K. The horizontal dashed line at 100 MPa represents the ``critical'' stress $\sigma_{c}$ at which the magnetoelastic anisotropy becomes comparable to the magnetocrystalline anisotropy in single-crystalline bulk fcc Ni with an anisotropy constant of $K_1$ = - 5 kJ/m$^3$ \cite{Cullity_2008}. This demonstrates that the strain-induced anisotropy associated with the SPT in V$_2$O$_3$ is comparable to or can even dominate the intrinsic crystal anisotropy in a ``textured'' ferromagnetic thin film.\\ 
\begin{figure}[h]
   \includegraphics[scale = 0.45, trim= 0cm 0cm -1.5cm 0cm, clip=true]{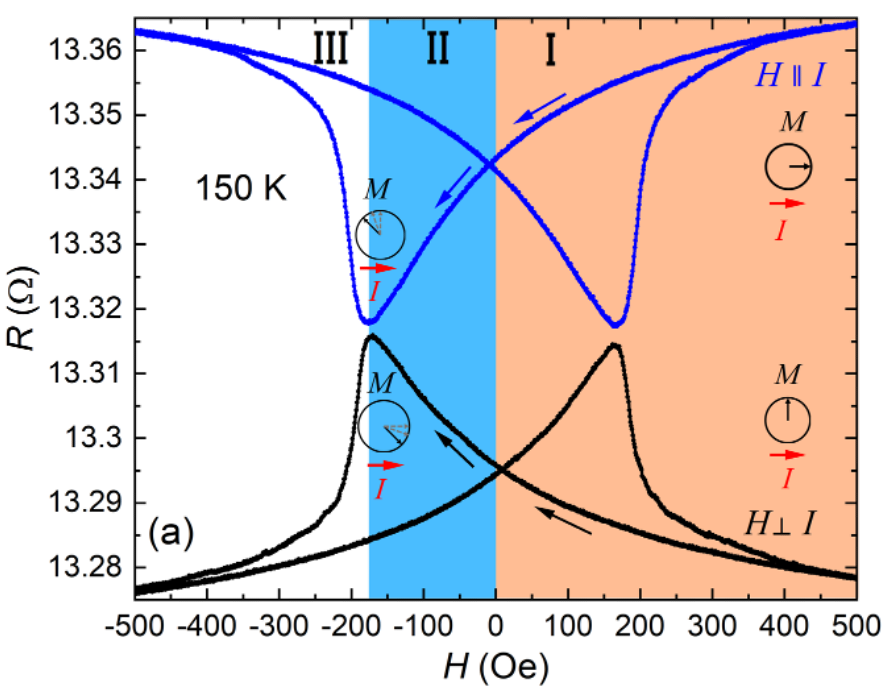}
    \includegraphics[scale = 0.3, trim= 1cm 3.2cm 0cm 2cm, clip=true]{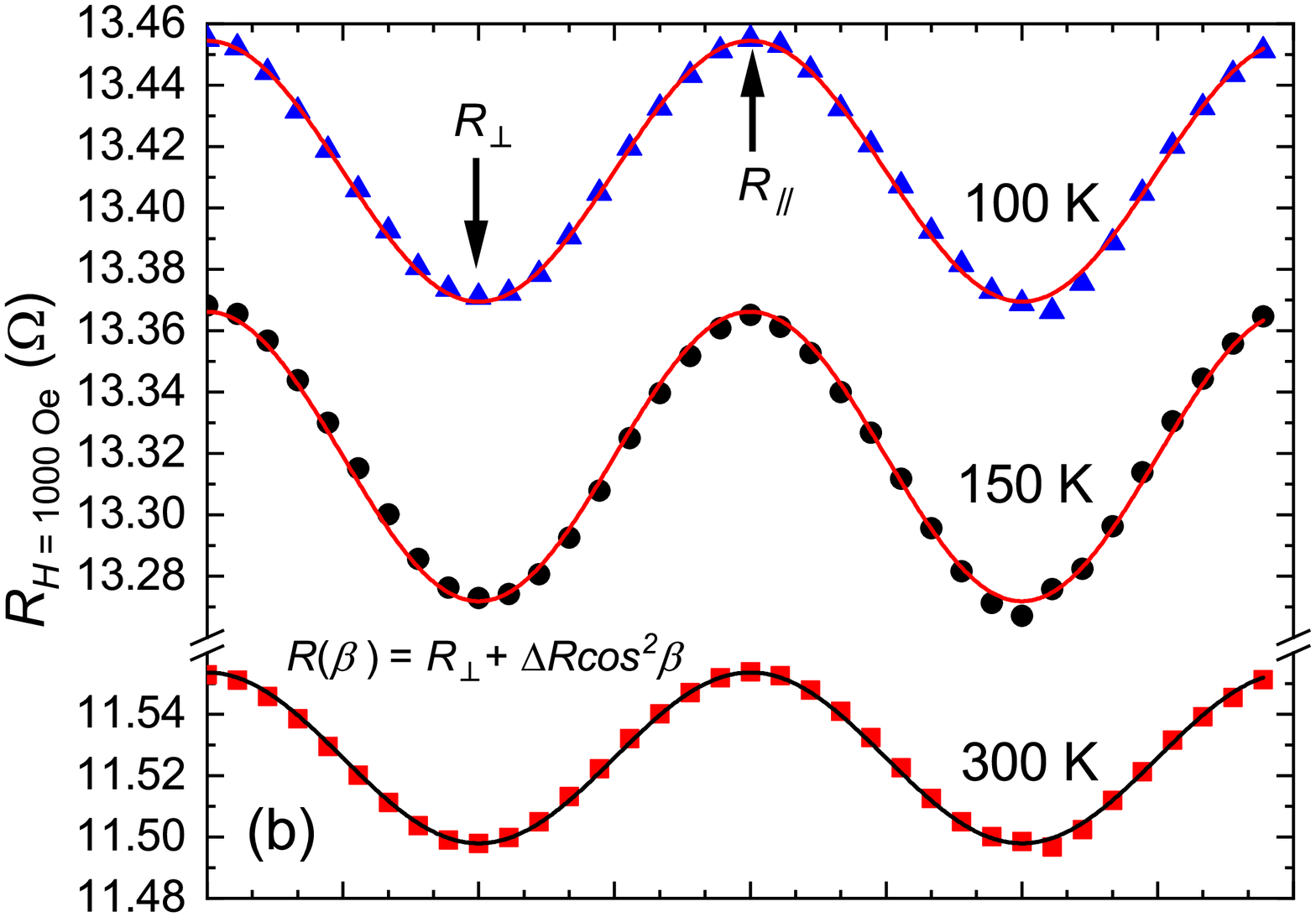}
     \includegraphics[scale = 0.3, trim= 1cm 1cm 0.25cm 2.3cm, clip=true]{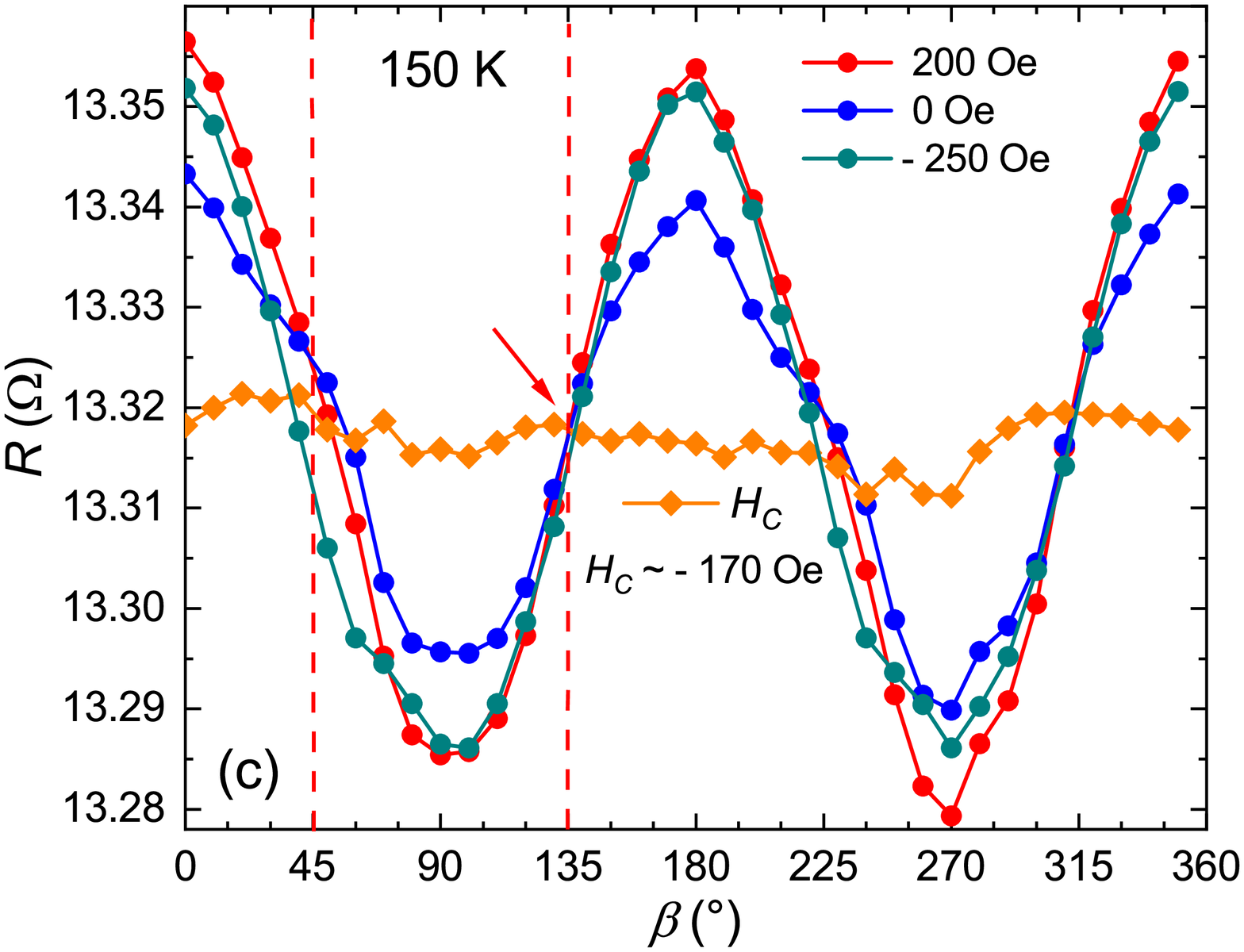}
    \vspace{-0.25 cm}
\caption{\label{fig6} (Color online) (a) The longitudinal and transverse magnetoresistance (MR) for the \textit{tS2} sample at 150 K. The blue (black) curves show the MR when the applied magnetic field is parallel $H_{\parallel}$ (or perpendicular $H_{\perp}$) to the direction of the electric current $I$. (b) The angular dependence of the electrical resistance $R(\beta)$ in strong field ($H$ = 1000 Oe) at 300, 150, and 100 K. The solid lines are fits of the function $R(\beta)$ = $R_{\perp}$ + $\Delta$$R$$cos^2\beta$ to the MR data where $\Delta$$R$ = $R_{\parallel}$ $-$ $R_{\perp}$ \cite{McGuire_1975}. (c) The angular dependence of the electrical resistance at various field at 150 K. The red arrow indicates there is an equality between $R$ at coercivity $H_c$ and $R$ in a strong magnetic field $H$ when directed at $\beta$ = 45\degree, 135\degree, 225\degree, and 315\degree.} 
\end{figure}
\indent Measurements of the anisotropic magnetoresistance (AMR) can be used to detect the strain-induced changes to the magnetization $M$  resulting from in-plane tensile stress \cite{Asai_2016}. Here, in combination with the FMR, measurements of AMR were used to better understand the effect of tensile strain on the magnetic anisotropy in the Ni layer.  The AMR measurements were performed on the  \textit{tS2} sample upon warming at 100, 150, and 300 K. The magnetic field $H$  was applied parallel to the Ni film surface and the angle between $H$ and the electric current $I$ was varied in intervals of 10\degree, where the initial angle of 0\degree~corresponds to alignment of the magnetic field $H$  with the electric current $I$. (See Fig.~\ref{fig1} in the Experimental section for a description of the geometry of the AMR measurement.) In this paper, we obtained the magnetoresistance (MR) by measuring the electrical resistance $R$ in a dc magnetic field $H$ applied at a given angle relative to the direction of electric current $I$. The AMR is the difference between the longitudinal electrical resistance $R_{\parallel}$ and the transverse electrical resistance $R_{\perp}$: $\Delta$$R$ = $R_{\parallel}$ - $R_{\perp}$. Figure~\ref{fig6}(a) displays the longitudinal and transverse magnetoresistance (MR) curves for the  \textit{tS2} sample between $H$ = $-$500 Oe and 500 Oe. At strong magnetic field $H$, the domains align parallel to $H$ such that $M$${\parallel}$$H$. The magnitude of the longitudinal resistance $R_{\parallel}$ (blue curve for $H_{\parallel}$) is consistently greater than the transverse resistance $R_{\perp}$ (black curve for $H_{\perp}$). This is a characteristic feature of polycrystalline samples of ferromagnetic materials, known as the AMR effect \cite{Bozorth_1946, McGuire_1975}.  The longitudinal and transverse MR at 100 K and 300 K exhibit the same effect and are included in Fig.~\ref{figS6} of the Supplemental Material.\\ 
\indent The electrical resistance $R$ is a measure of the spin-dependent scattering of the conduction electrons (s electrons) from the magnetic domains \cite{McGuire_1975, Coey_2010a}. The scattering of the current-carrying $s$ electrons from the localized $d$ electrons is largest (smallest) when the electrons move parallel (perpendicular) to the magnetization $M$ \cite{Smit_1951, McGuire_1975, Coey_2010a}. This yields the observed difference $R_{\parallel}$ $>$ $R_{\perp}$ as displayed in Fig.~\ref{fig6}(a). As the applied magnetic field is reduced from 1000 Oe to 0 Oe in region \RomanNumeralCaps{1} and then reversed to a coercive field $H_c$ in region \RomanNumeralCaps{2}, the magnetic domains reconfigure so that the magnetization $M$ rotates toward the EA along the $\phi$ = 0\degree~direction. At the coercive field $H_c$, the magnetization $M$ has rotated from its original strong field direction toward the EA and there is a corresponding decrease (or increase) in the MR toward a minimum (or maximum). The values of $R_{\parallel}$ and $R_{\perp}$ converge to nearly the same value at the coercive field $H_c$. This happens at a coercive field of roughly $H_c$ = $\pm$50 Oe at 300 K, and at $H_c$ = $\pm$170 Oe at 150 K and $H_c$ = $\pm$190 Oe at 100 K. 
Figure~\ref{fig6}(b) illustrates the typical cos$^2$ dependence of the electrical resistance $R$ of a polycrystalline sample of a ferromagnetic material at sufficiently large $H$. The solid lines are fits of the function $R(\beta)$ = $R_{\perp} + \Delta$$R$$cos^2\beta$ to the MR data, where $\Delta$$R$ = $R_{\parallel} - R_{\perp}$. This expression for $R(\beta)$ describes the typical angular dependence observed for the electrical resistance in a thin-film sample of a polycrystalline ferromagnetic material \cite{McGuire_1975}. The angular dependence of the magnetoresistance MR at various values of applied magnetic field $H$, ranging from the saturation field to the coercive field $H_c$, is shown at 150 K in Fig.~\ref{fig6}(c). It is important to emphasize that the cos$^2\beta$  dependence of $R(\beta)$ vanishes at the coercive field $H_c$ = $-$170 Oe. The magnitude of the electrical resistance $R$ $\approx$ 13.32 $\Omega$ is nearly constant at $H_c$ and is equal in magnitude to the electrical resistance $R$ when in strong magnetic field $H$ that is directed along $\beta$ = 45\degree, 135\degree, 225\degree, and 315\degree.   Similar behavior is observed in the MR for the  \textit{tS2} sample above and below the transition temperature $T_C$, where the values of the electrical resistance at $H_c$ are $R$ $\approx$ 11.53 $\Omega$  and 13.41 $\Omega$ at 300 K and 100 K, respectively. (See Fig.~\ref{figS7} in the Supplemental Material.) At these four orientations of the magnetic field ($\beta$ = 45\degree, 135\degree, 225\degree, and 315\degree), the magnetic domains would have angular projections of 45\degree~to the direction of the electric current $I$. In order for the magnitude of $R$ at the coercive field $H_c$ to be equal to the specific values of $R$(45\degree), $R$(135\degree), $R$(225\degree), and $R$(315\degree) in strong field as shown in Fig.~\ref{fig6}(c), the scattering mechanism for the itinerant electrons is expected to be similar in each case. This suggests that the domains will align to their default orientation (in a relatively weak field near $H_c$) that is at a 45\degree~projection relative to the direction of electric current $I$.\section{Discussion}
\label{Discussion}
\begin{figure}[t]
   \includegraphics[scale = 0.4, trim= 3.5cm 5cm 3.5cm 3.5cm, clip=true]{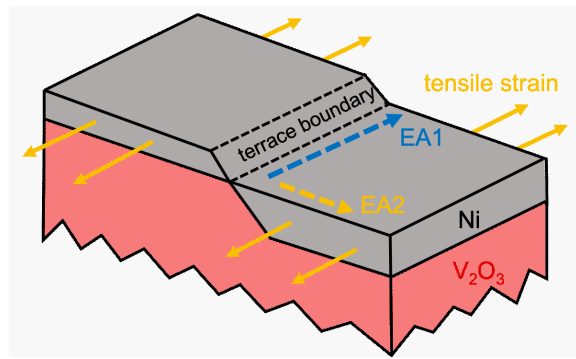}
\caption{\label{fig7} (Color online) A schematic (not to scale) of the growth-induced microstructure of the V$_2$O$_3$/Ni bilayer \cite{Gilbert_2017}. The tensile strain (solid yellow arrows) is effectively uniaxial along the terrace boundary. The growth-induced easy axis EA1 (dashed blue arrow) is parallel to the terrace boundary while the strain-induced secondary easy axis EA2 (dashed yellow arrow) is perpendicular to the terrace boundary.} 
\end{figure}

\begin{figure}[t]
   \includegraphics[scale = 0.35, trim= 1.0cm 3cm -2cm 2.5cm, clip=true]{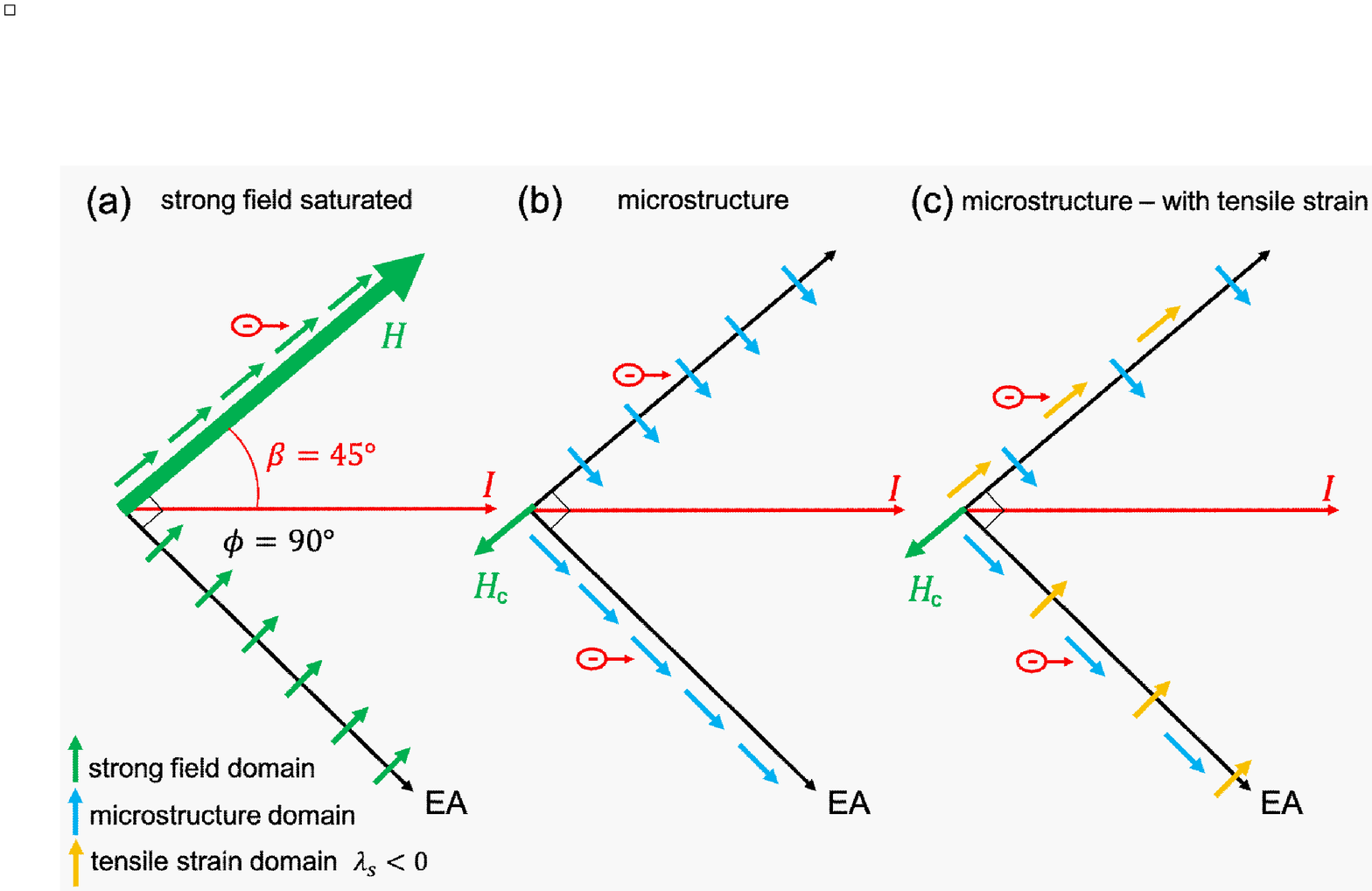}
\caption{\label{fig8} (Color online) A schematic of the domains at field saturation (green arrows) compared to the domains aligned with the terraced microstructure (blue arrows) and the domains resulting from the uniaxial strain along the easy axis (yellow arrows). The yellow arrows are perpendicular to the primary easy axis owing to the negative magnetostriction coefficient $\lambda_{s}$ of Ni. The scattering of electrons (red) from the domains is the same in each configuration. (Note: the small coercive field $H_c$ applied at all angles is shown here at 45\degree~as an example.)} 
\end{figure}

A comparison of the structural and magnetic properties of two V$_2$O$_3$/Ni bilayer samples at high and low temperature allows for an indirect determination of the tensile strain in the Ni layer. Two different sources of strain in the Ni layer are identified. The strongest strain is produced by the structural phase transition (SPT) from the high-temperature (HT) rhombohedral phase to the low-temperature (LT) monoclinic phase in V$_2$O$_3$ at $T_C$ = 160 K. Upon cooling through the SPT into the monoclinic phase, the volume of the unit cell expands approximately 1.4\%, during which there is an increase in the $a$ lattice parameter from 2.87 to 2.91 \AA~\cite{Singer_2018}. This expansion produces an in-plane tensile strain in the proximal Ni layer. A weaker strain in the Ni layer was observed at HT only in sample $ts2$.\\
\indent We argue that the HT biaxial anisotropy observed in the FMR for sample $ts2$ only (see Fig.~\ref{fig5} (c)) is a proximity-induced strain in the Ni layer caused by a distorted rhombohedral phase of V$_2$O$_3$. The distortion of the V$_2$O$_3$ layer is caused by the misalignment of the V$_2$O$_3$ and Al$_2$O$_3$ (012) planes and is ``frozen in'' at room temperature during growth (see Fig.~\ref{fig2}) \cite{McLeod_2017, Kalcheim_2019}. This distortion in 100 nm films of rhombohedral phase V$_2$O$_3$ at room temperature can be comparable in magnitude to the strain produced during the rhombohedral to monoclinic SPT at $T_C$ = 160 K \cite{Kalcheim_2019}. These two different sources of strain (SPT-induced and growth-induced) produce the same biaxial anisotropy, suggesting that the nature of the tensile strain produced in the Ni layer is the same in both cases. The emergence of a strain-induced secondary easy axis (EA) in the magnetic anisotropy serves to identify the direction of tensile strain in the Ni layer. This follows from the negative magnetostriction of ferromagnetic Ni where the strain-induced secondary EA is perpendicular to direction of tensile strain in the Ni layer. Here, we argue that the tensile strain is effectively uniaxial and is directed along the primary EA at $\phi =$ 0\degree, which is parallel to the terrace boundaries in the microstructure of the V$_2$O$_3$/Ni bilayer as discussed previously in Ref.~\onlinecite{Gilbert_2017}.\\ 
\indent Direct observation of the strain-induced reversal of magnetic domains in Ni across the SPT in V$_2$O$_3$ was recently reported in Ref.~\onlinecite{Valmianski_2021}. First order reversal curve (FORC) measurements reported in Ref.~\onlinecite{Gilbert_2017} also indicate that changes to the magnetization of the V$_2$O$_3$/Ni bilayer across the SPT in V$_2$O$_3$ are caused by ``irreversible'' switching mechanisms attributed to an increase in strain-induced pinning \cite{Davies_2004}. Both the reversible and irreversible domain switching occurs in a magneto-structural landscape, where the individual magnetic domains are confined to the Ni regions bounded by the ``rips'' in the microstructure. Previous reports suggest that the ruptures in the Ni layer occur where there is a particularly large stress at the terrace boundaries \cite{Gilbert_2017, McLeod_2017, Abadias_2018}. (The flexural strength of Ni, also known as the bend strength or modulus of rupture, can take on values as small as $\sigma_r$ = 70 MPa and has a maximum value on the order of $\sigma_r$ =  935 MPa \cite{IMMA_1997}.) The directionality of the ruptures in the Ni film suggests the film has reached an elastic limit in the direction perpendicular to the terrace boundaries \cite{Gilbert_2017, McLeod_2017}. Here, we present a scenario in which the tensile strain is mechanically coupled to the ruptured microstructure of the V$_2$O$_3$/Ni bilayer. The strain is channeled through the Ni regions that are confined by the rips and is parallel to the terraced-induced EA at $\phi =$ 0\degree. A schematic of the tensile strain in relation to the terrace boundaries is displayed in Fig.~\ref{fig7}. \\
\indent The tensile strain competes with the local anisotropy (produced by the terraces, defects, inclusions, grain boundaries, etc.) to produce a secondary EA as observed in the FMR shown in Fig.~\ref{fig3}. Owing to the negative magnetostriction coefficient $\lambda_{s}$ for Ni, the secondary EA emerges in a direction that is orthogonal to the tensile strain direction \cite{Asai_2016, Bozorth_1946}. 
This multi-domain landscape is presented in Fig.~\ref{fig8}, where the behavior of the individual magnetic domains is governed by the competing influence of (a) an applied magnetic field, (b) the terraces in the microstructure, and (c) the tensile strain. The schematic in Fig.~\ref{fig8}(a) displays the domain configuration (small green arrows) in a sufficiently strong magnetic field $H$ (large green arrow) that is directed at $\beta$ = 45\degree~with respect to the direction of electric current. The spin-dependent scattering occurs with the magnetic domains saturated in an orientation at 45\degree~relative the motion of the itinerant electrons (red symbols). In comparison, the schematic representations in Figs.~\ref{fig8}(b) and (c) present the default configuration of magnetic domains in a relatively small coercive field $H_c$ (green arrow). In the absence of tensile strain (Fig.~\ref{fig8}(b)), the terraces in the surface morphology determine the preferred orientation of magnetic domains (blue arrows), which align parallel to the terrace boundaries \cite{Chuang_1994, Albrecht_1992, Gilbert_2017}. This spontaneous orientation of domains is reflected in the growth-induced uniaxial magnetic anisotropy that exists at high temperature and persists through the first-order SPT to lower temperatures. The persistence of the primary EA at $\phi =$ 0\degree~is evident from the temperature dependent FMR shown in Fig.~\ref{fig3}. In the presence of a uniaxial tensile strain (Fig.~\ref{fig8}(c)), the negative Ni magnetostriction $\lambda_{s}$ forces the domains (yellow arrows) to orient perpendicular to the strain direction. This stress anisotropy competes with the local microstructure-induced anisotropy to produce an emergent biaxial anisotropy with a secondary EA at $\phi =$90\degree~(see FMR in Fig.~\ref{fig3}).\\ 
\indent The domain configuration described above suggests an equivalence between the spin-dependent scattering at the coercive field $H_c$ (Fig.~\ref{fig8}(b) and Fig.~\ref{fig8}(c)) and in strong magnetic field oriented at $\beta$ = 45\degree~(Fig.~\ref{fig8}(a)) with respect to the electric current (red arrow). 
This is reflected in the AMR, where the electrical resistances $R$ are equal at the coercive field $H_c$ and in a strong magnetic field $H$  at $\beta$ = 45\degree, 135\degree, 225\degree, and 315\degree. This is indicated by the red arrow in Fig.~\ref{fig6}(c) and in Fig.~\ref{figS7} in the Supplemental Material. Owing to the experimental setup and geometry of the electrical resistance measurement in this study, the direction of the electric current bisects the 90\degree~angle between the primary and secondary magnetic easy axes. (The electric current is fixed at 45\degree~relative to EA at $\phi =$ 0\degree.)  This symmetry results in an equivalence between the spin-dependent scattering from the primary axis in the absence of strain and from the secondary easy axis in the presence of strain. Hence, the scattering angle is the same before (Fig.~\ref{fig8}(b)) and after (Fig.~\ref{fig8} (c)) the strain-induced 90\degree~reversal of the magnetic domains. 
This equivalence in the spin-dependent scattering angle both above and below the SPT supports a scenario for a strain-induced 90\degree~reversal of magnetic domains within the Ni layer that is consistent with the emergence of a secondary easy axis observed in the FMR.
\section{Conclusions}
\label{Conclusions}
We have studied the reversible and controllable changes to the in-plane magnetic anisotropy in V$_2$O$_3$/Ni bilayers deposited on $r$-cut sapphire substrates. The magnetic anisotropy determined from the ferromagnetic resonance is described as a superposition of uniaxial and biaxial anisotropy. At high temperature, the anisotropy is predominantly uniaxial with a single easy axis. Upon cooling through the structural change in V$_2$O$_3$ at 160 K, a secondary strain-induced easy axis emerges in the Ni. This emergent biaxial anisotropy is caused by the expansion of the lattice during the rhombohedral to monoclinic transition in the underlying V$_2$O$_3$ layer. This biaxial anisotropy that emerges upon cooling through the structural phase transition in V$_2$O$_3$ is reversible in both V$_2$O$_3$/Ni samples studied here. A similar but weaker biaxial anisotropy at high temperature is observed in one of the V$_2$O$_3$/Ni bilayer samples. This biaxial anisotropy is also stress-induced, where the tensile strain in the Ni layer is caused by the ``distortion'' of the underlying V$_2$O$_3$ layer due to misalignment of the V$_2$O$_3$ layer with the sapphire substrate. This strain-induced biaxial anisotropy is not reversible and is ``frozen into'' the V$_2$O$_3$/Ni bilayer during growth.\\ 
\indent From the X-ray crystallography, ferromagnetic resonance, and anisotropic magnetoresistance, we have indirectly determined the direction of the tensile strain in two V$_2$O$_3$/Ni bilayer samples grown on \text{r}-cut sapphire substrates. Moreover, the tensile strain producing the changes in the magnetic anisotropy is coupled to the terraced microstructure in the V$_2$O$_3$/Ni bilayer. Specifically, there is a preferred direction to the rupturing of the Ni layer that is connected to the terrace boundaries. This directionality in the rupturing of the Ni layer forces the in-plane strain in a direction parallel to the terrace boundaries. Hence, the in-plane strain in the Ni layer is effectively a uniaxial strain that is directed along the regions of Ni that exist between the terrace-induced rips. Owing to the negative magnetostriction of Ni, this uniaxial strain along the terrace boundaries accounts for the emergence of a secondary easy axis that is perpendicular to the primary magnetic easy axis along terrace boundaries.\\ 
\indent We suggest that a reversible and reproducible tensile strain could be used to switch the in-plane magnetic anisotropy in ferromagnetic thin films (in both polycrystalline and quasi-textured layers). This relies initially on the growth-induced uniaxial anisotropy, caused by terraces in the microstructure, which overcomes the magnetocrystalline anisotropy as observed in the case of Ni. An additional ``switchable'' anisotropy could then be applied externally by the incorporation of the ferromagnetic layer into an active heterostructure. In the V$_2$O$_3$/Ni bilayer system investigated here, we take advantage of the structural change in the proximal V$_2$O$_3$ layer. In this case, the stress anisotropy produced by the structural transition in the V$_2$O$_3$ layer causes an anisotropy comparable or larger than the intrinsic crystalline anisotropy of bulk Ni. 
\begin{acknowledgments}
The research was supported by the U.S. Department of Energy (DOE), Office of Science, Basic Energy Sciences (BES), Materials Sciences and Engineering Division under Award \# DE-FG02-87ER-45332.
\end{acknowledgments}
\bibliography{V2O3_Ni}

\providecommand{\noopsort}[1]{}\providecommand{\singleletter}[1]{#1}%
\begin{thebibliography}{34}%
\makeatletter
\providecommand \@ifxundefined [1]{%
 \@ifx{#1\undefined}
}%
\providecommand \@ifnum [1]{%
 \ifnum #1\expandafter \@firstoftwo
 \else \expandafter \@secondoftwo
 \fi
}%
\providecommand \@ifx [1]{%
 \ifx #1\expandafter \@firstoftwo
 \else \expandafter \@secondoftwo
 \fi
}%
\providecommand \natexlab [1]{#1}%
\providecommand \enquote  [1]{``#1''}%
\providecommand \bibnamefont  [1]{#1}%
\providecommand \bibfnamefont [1]{#1}%
\providecommand \citenamefont [1]{#1}%
\providecommand \href@noop [0]{\@secondoftwo}%
\providecommand \href [0]{\begingroup \@sanitize@url \@href}%
\providecommand \@href[1]{\@@startlink{#1}\@@href}%
\providecommand \@@href[1]{\endgroup#1\@@endlink}%
\providecommand \@sanitize@url [0]{\catcode `\\12\catcode `\$12\catcode
  `\&12\catcode `\#12\catcode `\^12\catcode `\_12\catcode `\%12\relax}%
\providecommand \@@startlink[1]{}%
\providecommand \@@endlink[0]{}%
\providecommand \url  [0]{\begingroup\@sanitize@url \@url }%
\providecommand \@url [1]{\endgroup\@href {#1}{\urlprefix }}%
\providecommand \urlprefix  [0]{URL }%
\providecommand \Eprint [0]{\href }%
\providecommand \doibase [0]{https://doi.org/}%
\providecommand \selectlanguage [0]{\@gobble}%
\providecommand \bibinfo  [0]{\@secondoftwo}%
\providecommand \bibfield  [0]{\@secondoftwo}%
\providecommand \translation [1]{[#1]}%
\providecommand \BibitemOpen [0]{}%
\providecommand \bibitemStop [0]{}%
\providecommand \bibitemNoStop [0]{.\EOS\space}%
\providecommand \EOS [0]{\spacefactor3000\relax}%
\providecommand \BibitemShut  [1]{\csname bibitem#1\endcsname}%
\let\auto@bib@innerbib\@empty
\bibitem [{\citenamefont {Albrecht}\ \emph {et~al.}(1992)\citenamefont
  {Albrecht}, \citenamefont {Furubayashi}, \citenamefont {Przybylski},
  \citenamefont {Korecki},\ and\ \citenamefont {Gradman}}]{Albrecht_1992}%
  \BibitemOpen
  \bibfield  {author} {\bibinfo {author} {\bibfnamefont {M.}~\bibnamefont
  {Albrecht}}, \bibinfo {author} {\bibfnamefont {T.}~\bibnamefont
  {Furubayashi}}, \bibinfo {author} {\bibfnamefont {M.}~\bibnamefont
  {Przybylski}}, \bibinfo {author} {\bibfnamefont {J.}~\bibnamefont
  {Korecki}},\ and\ \bibinfo {author} {\bibfnamefont {U.}~\bibnamefont
  {Gradman}},\ }\bibfield  {title} {\bibinfo {title} {Magnetic step
  anisotropies},\ }\href
  {https://doi.org/https://doi.org/10.1016/0304-8853(92)91269-Y} {\bibfield
  {journal} {\bibinfo  {journal} {Journal of Magnetism and Magnetic Materials}\
  }\textbf {\bibinfo {volume} {113}},\ \bibinfo {pages} {207} (\bibinfo {year}
  {1992})}\BibitemShut {NoStop}%
\bibitem [{\citenamefont {Metoki}\ \emph {et~al.}(1993)\citenamefont {Metoki},
  \citenamefont {Zeidler}, \citenamefont {Stierle}, \citenamefont {Bröhl},\
  and\ \citenamefont {Zabel}}]{Metoki_1993}%
  \BibitemOpen
  \bibfield  {author} {\bibinfo {author} {\bibfnamefont {N.}~\bibnamefont
  {Metoki}}, \bibinfo {author} {\bibfnamefont {T.}~\bibnamefont {Zeidler}},
  \bibinfo {author} {\bibfnamefont {A.}~\bibnamefont {Stierle}}, \bibinfo
  {author} {\bibfnamefont {K.}~\bibnamefont {Bröhl}},\ and\ \bibinfo {author}
  {\bibfnamefont {H.}~\bibnamefont {Zabel}},\ }\bibfield  {title} {\bibinfo
  {title} {Uniaxial magnetic anisotropy of co films on sapphire},\ }\href
  {https://doi.org/https://doi.org/10.1016/0304-8853(93)90157-W} {\bibfield
  {journal} {\bibinfo  {journal} {Journal of Magnetism and Magnetic Materials}\
  }\textbf {\bibinfo {volume} {118}},\ \bibinfo {pages} {57} (\bibinfo {year}
  {1993})}\BibitemShut {NoStop}%
\bibitem [{\citenamefont {Chuang}\ \emph {et~al.}(1994)\citenamefont {Chuang},
  \citenamefont {Ballentine},\ and\ \citenamefont {O'Handley}}]{Chuang_1994}%
  \BibitemOpen
  \bibfield  {author} {\bibinfo {author} {\bibfnamefont {D.~S.}\ \bibnamefont
  {Chuang}}, \bibinfo {author} {\bibfnamefont {C.~A.}\ \bibnamefont
  {Ballentine}},\ and\ \bibinfo {author} {\bibfnamefont {R.~C.}\ \bibnamefont
  {O'Handley}},\ }\bibfield  {title} {\bibinfo {title} {Surface and step
  magnetic anisotropy},\ }\href {https://doi.org/10.1103/PhysRevB.49.15084}
  {\bibfield  {journal} {\bibinfo  {journal} {Phys. Rev. B}\ }\textbf {\bibinfo
  {volume} {49}},\ \bibinfo {pages} {15084} (\bibinfo {year}
  {1994})}\BibitemShut {NoStop}%
\bibitem [{\citenamefont {Kawakami}\ \emph {et~al.}(1996)\citenamefont
  {Kawakami}, \citenamefont {Escorcia-Aparicio},\ and\ \citenamefont
  {Qiu}}]{Kawakami_1996}%
  \BibitemOpen
  \bibfield  {author} {\bibinfo {author} {\bibfnamefont {R.~K.}\ \bibnamefont
  {Kawakami}}, \bibinfo {author} {\bibfnamefont {E.~J.}\ \bibnamefont
  {Escorcia-Aparicio}},\ and\ \bibinfo {author} {\bibfnamefont {Z.~Q.}\
  \bibnamefont {Qiu}},\ }\bibfield  {title} {\bibinfo {title} {Symmetry-induced
  magnetic anisotropy in {Fe} films grown on stepped {Ag}(001)},\ }\href
  {https://doi.org/10.1103/PhysRevLett.77.2570} {\bibfield  {journal} {\bibinfo
   {journal} {Phys. Rev. Lett.}\ }\textbf {\bibinfo {volume} {77}},\ \bibinfo
  {pages} {2570} (\bibinfo {year} {1996})}\BibitemShut {NoStop}%
\bibitem [{\citenamefont {Liu}\ \emph {et~al.}(2013)\citenamefont {Liu},
  \citenamefont {He}, \citenamefont {Wu}, \citenamefont {Ye}, \citenamefont
  {Zhang}, \citenamefont {Yang},\ and\ \citenamefont {Cheng}}]{Liu_2013}%
  \BibitemOpen
  \bibfield  {author} {\bibinfo {author} {\bibfnamefont {H.-L.}\ \bibnamefont
  {Liu}}, \bibinfo {author} {\bibfnamefont {W.}~\bibnamefont {He}}, \bibinfo
  {author} {\bibfnamefont {Q.}~\bibnamefont {Wu}}, \bibinfo {author}
  {\bibfnamefont {J.}~\bibnamefont {Ye}}, \bibinfo {author} {\bibfnamefont
  {X.-Q.}\ \bibnamefont {Zhang}}, \bibinfo {author} {\bibfnamefont {H.-T.}\
  \bibnamefont {Yang}},\ and\ \bibinfo {author} {\bibfnamefont {Z.-H.}\
  \bibnamefont {Cheng}},\ }\bibfield  {title} {\bibinfo {title} {Magnetic
  anisotropy of ultrathin {Fe} films grown on vicinal {Si} (111)},\ }\href
  {https://doi.org/10.1063/1.4809664} {\bibfield  {journal} {\bibinfo
  {journal} {AIP Advances}\ }\textbf {\bibinfo {volume} {3}},\ \bibinfo {pages}
  {062101} (\bibinfo {year} {2013})}\BibitemShut {NoStop}%
\bibitem [{\citenamefont {Davydenko}\ \emph {et~al.}(2014)\citenamefont
  {Davydenko}, \citenamefont {Kozlov},\ and\ \citenamefont
  {Chebotkevich}}]{Davydenko_2014}%
  \BibitemOpen
  \bibfield  {author} {\bibinfo {author} {\bibfnamefont {A.~V.}\ \bibnamefont
  {Davydenko}}, \bibinfo {author} {\bibfnamefont {A.~G.}\ \bibnamefont
  {Kozlov}},\ and\ \bibinfo {author} {\bibfnamefont {L.~A.}\ \bibnamefont
  {Chebotkevich}},\ }\bibfield  {title} {\bibinfo {title} {Spatial modulation
  of in-plane magnetic anisotropy in epitaxial {Co}(111) films grown on
  macrostep-bunched {Si}(111)},\ }\href {https://doi.org/10.1063/1.4897536}
  {\bibfield  {journal} {\bibinfo  {journal} {Journal of Applied Physics}\
  }\textbf {\bibinfo {volume} {116}},\ \bibinfo {pages} {143901} (\bibinfo
  {year} {2014})}\BibitemShut {NoStop}%
\bibitem [{\citenamefont {Kittel}(1949)}]{Kittel_1949}%
  \BibitemOpen
  \bibfield  {author} {\bibinfo {author} {\bibfnamefont {C.}~\bibnamefont
  {Kittel}},\ }\bibfield  {title} {\bibinfo {title} {{Physical Theory of
  Ferromagnetic Domains}},\ }\href {https://doi.org/10.1103/RevModPhys.21.541}
  {\bibfield  {journal} {\bibinfo  {journal} {Rev. Mod. Phys.}\ }\textbf
  {\bibinfo {volume} {21}},\ \bibinfo {pages} {541} (\bibinfo {year}
  {1949})}\BibitemShut {NoStop}%
\bibitem [{\citenamefont {Lee}(1955)}]{Lee_1955}%
  \BibitemOpen
  \bibfield  {author} {\bibinfo {author} {\bibfnamefont {E.~W.}\ \bibnamefont
  {Lee}},\ }\bibfield  {title} {\bibinfo {title} {{Magnetostriction and
  Magnetomechanical Effects}},\ }\href
  {https://doi.org/10.1088/0034-4885/18/1/305} {\bibfield  {journal} {\bibinfo
  {journal} {Reports on Progress in Physics}\ }\textbf {\bibinfo {volume}
  {18}},\ \bibinfo {pages} {184} (\bibinfo {year} {1955})}\BibitemShut
  {NoStop}%
\bibitem [{\citenamefont {Liu}\ \emph {et~al.}(2014)\citenamefont {Liu},
  \citenamefont {Wang}, \citenamefont {Zhan}, \citenamefont {Tang},
  \citenamefont {Yang}, \citenamefont {Zuo}, \citenamefont {Zhang},
  \citenamefont {Xie}, \citenamefont {Zhu}, \citenamefont {Chen}, \citenamefont
  {Wang},\ and\ \citenamefont {Li}}]{Liu_2014}%
  \BibitemOpen
  \bibfield  {author} {\bibinfo {author} {\bibfnamefont {Y.}~\bibnamefont
  {Liu}}, \bibinfo {author} {\bibfnamefont {B.}~\bibnamefont {Wang}}, \bibinfo
  {author} {\bibfnamefont {Q.}~\bibnamefont {Zhan}}, \bibinfo {author}
  {\bibfnamefont {Z.}~\bibnamefont {Tang}}, \bibinfo {author} {\bibfnamefont
  {H.}~\bibnamefont {Yang}}, \bibinfo {author} {\bibfnamefont {Z.}~\bibnamefont
  {Zuo}}, \bibinfo {author} {\bibfnamefont {X.}~\bibnamefont {Zhang}}, \bibinfo
  {author} {\bibfnamefont {Y.}~\bibnamefont {Xie}}, \bibinfo {author}
  {\bibfnamefont {X.}~\bibnamefont {Zhu}}, \bibinfo {author} {\bibfnamefont
  {B.}~\bibnamefont {Chen}}, \bibinfo {author} {\bibfnamefont {J.}~\bibnamefont
  {Wang}},\ and\ \bibinfo {author} {\bibfnamefont {R.-W.}\ \bibnamefont {Li}},\
  }\bibfield  {title} {\bibinfo {title} {{Positive temperature coefficient of
  magnetic anisotropy in polyvinylidene fluoride ({PVDF})-based magnetic
  composites}},\ }\href {https://doi.org/10.1038/srep06615} {\bibfield
  {journal} {\bibinfo  {journal} {Scientific Reports}\ }\textbf {\bibinfo
  {volume} {4}},\ \bibinfo {pages} {6615} (\bibinfo {year} {2014})}\BibitemShut
  {NoStop}%
\bibitem [{\citenamefont {Gilbert}\ \emph {et~al.}(2017)\citenamefont
  {Gilbert}, \citenamefont {Ram\'{i}rez}, \citenamefont {Saerbeck},
  \citenamefont {Tralstoy}, \citenamefont {Schuller}, \citenamefont {Liu},\
  and\ \citenamefont {de~la Venta}}]{Gilbert_2017}%
  \BibitemOpen
  \bibfield  {author} {\bibinfo {author} {\bibfnamefont {D.~A.}\ \bibnamefont
  {Gilbert}}, \bibinfo {author} {\bibfnamefont {J.~G.}\ \bibnamefont
  {Ram\'{i}rez}}, \bibinfo {author} {\bibfnamefont {T.}~\bibnamefont
  {Saerbeck}}, \bibinfo {author} {\bibfnamefont {J.}~\bibnamefont {Tralstoy}},
  \bibinfo {author} {\bibfnamefont {I.~K.}\ \bibnamefont {Schuller}}, \bibinfo
  {author} {\bibfnamefont {K.}~\bibnamefont {Liu}},\ and\ \bibinfo {author}
  {\bibfnamefont {J.}~\bibnamefont {de~la Venta}},\ }\bibfield  {title}
  {\bibinfo {title} {{Growth-Induced In-Plane Uniaxial Anisotropy in V$_2$O
  $_3$ Films}},\ }\href {https://doi.org/10.1038/s41598-017-12690-z} {\bibfield
   {journal} {\bibinfo  {journal} {Scientific Reports}\ }\textbf {\bibinfo
  {volume} {7}},\ \bibinfo {pages} {13471} (\bibinfo {year}
  {2017})}\BibitemShut {NoStop}%
\bibitem [{\citenamefont {de~la Venta}\ \emph {et~al.}(2013)\citenamefont
  {de~la Venta}, \citenamefont {Wang}, \citenamefont {Ramirez},\ and\
  \citenamefont {Schuller}}]{delaVenta_2013}%
  \BibitemOpen
  \bibfield  {author} {\bibinfo {author} {\bibfnamefont {J.}~\bibnamefont
  {de~la Venta}}, \bibinfo {author} {\bibfnamefont {S.}~\bibnamefont {Wang}},
  \bibinfo {author} {\bibfnamefont {J.~G.}\ \bibnamefont {Ramirez}},\ and\
  \bibinfo {author} {\bibfnamefont {I.~K.}\ \bibnamefont {Schuller}},\
  }\bibfield  {title} {\bibinfo {title} {Control of magnetism across metal to
  insulator transitions},\ }\href {https://doi.org/10.1063/1.4798293}
  {\bibfield  {journal} {\bibinfo  {journal} {Applied Physics Letters}\
  }\textbf {\bibinfo {volume} {102}},\ \bibinfo {pages} {122404} (\bibinfo
  {year} {2013})}\BibitemShut {NoStop}%
\bibitem [{\citenamefont {de~la Venta}\ \emph {et~al.}(2014)\citenamefont
  {de~la Venta}, \citenamefont {Wang}, \citenamefont {Saerbeck}, \citenamefont
  {Ramírez}, \citenamefont {Valmianski},\ and\ \citenamefont
  {Schuller}}]{delaVenta_2014}%
  \BibitemOpen
  \bibfield  {author} {\bibinfo {author} {\bibfnamefont {J.}~\bibnamefont
  {de~la Venta}}, \bibinfo {author} {\bibfnamefont {S.}~\bibnamefont {Wang}},
  \bibinfo {author} {\bibfnamefont {T.}~\bibnamefont {Saerbeck}}, \bibinfo
  {author} {\bibfnamefont {J.~G.}\ \bibnamefont {Ramírez}}, \bibinfo {author}
  {\bibfnamefont {I.}~\bibnamefont {Valmianski}},\ and\ \bibinfo {author}
  {\bibfnamefont {I.~K.}\ \bibnamefont {Schuller}},\ }\bibfield  {title}
  {\bibinfo {title} {{Coercivity enhancement in V$_2$O$_3$/Ni bilayers driven
  by nanoscale phase coexistence}},\ }\href {https://doi.org/10.1063/1.4865587}
  {\bibfield  {journal} {\bibinfo  {journal} {Applied Physics Letters}\
  }\textbf {\bibinfo {volume} {104}},\ \bibinfo {pages} {062410} (\bibinfo
  {year} {2014})}\BibitemShut {NoStop}%
\bibitem [{\citenamefont {Kalcheim}\ \emph {et~al.}(2019)\citenamefont
  {Kalcheim}, \citenamefont {Butakov}, \citenamefont {Vargas}, \citenamefont
  {Lee}, \citenamefont {del Valle}, \citenamefont {Trastoy}, \citenamefont
  {Salev}, \citenamefont {Schuller},\ and\ \citenamefont
  {Schuller}}]{Kalcheim_2019}%
  \BibitemOpen
  \bibfield  {author} {\bibinfo {author} {\bibfnamefont {Y.}~\bibnamefont
  {Kalcheim}}, \bibinfo {author} {\bibfnamefont {N.}~\bibnamefont {Butakov}},
  \bibinfo {author} {\bibfnamefont {N.~M.}\ \bibnamefont {Vargas}}, \bibinfo
  {author} {\bibfnamefont {M.-H.}\ \bibnamefont {Lee}}, \bibinfo {author}
  {\bibfnamefont {J.}~\bibnamefont {del Valle}}, \bibinfo {author}
  {\bibfnamefont {J.}~\bibnamefont {Trastoy}}, \bibinfo {author} {\bibfnamefont
  {P.}~\bibnamefont {Salev}}, \bibinfo {author} {\bibfnamefont
  {J.}~\bibnamefont {Schuller}},\ and\ \bibinfo {author} {\bibfnamefont
  {I.~K.}\ \bibnamefont {Schuller}},\ }\bibfield  {title} {\bibinfo {title}
  {{Robust Coupling between Structural and Electronic Transitions in a Mott
  Material}},\ }\href {https://doi.org/10.1103/PhysRevLett.122.057601}
  {\bibfield  {journal} {\bibinfo  {journal} {Phys. Rev. Lett.}\ }\textbf
  {\bibinfo {volume} {122}},\ \bibinfo {pages} {057601} (\bibinfo {year}
  {2019})}\BibitemShut {NoStop}%
\bibitem [{\citenamefont {Kosuge}(1967)}]{Kosuge_1967}%
  \BibitemOpen
  \bibfield  {author} {\bibinfo {author} {\bibfnamefont {K.}~\bibnamefont
  {Kosuge}},\ }\bibfield  {title} {\bibinfo {title} {{The phase diagram and
  phase transition of the V$_2$O$_3$-V$_2$O$_5$, system}},\ }\href
  {https://doi.org/https://doi.org/10.1016/0022-3697(67)90293-4} {\bibfield
  {journal} {\bibinfo  {journal} {Journal of Physics and Chemistry of Solids}\
  }\textbf {\bibinfo {volume} {28}},\ \bibinfo {pages} {1613} (\bibinfo {year}
  {1967})}\BibitemShut {NoStop}%
\bibitem [{\citenamefont {Dernier}\ and\ \citenamefont
  {Marezio}(1970)}]{Dernier_1970}%
  \BibitemOpen
  \bibfield  {author} {\bibinfo {author} {\bibfnamefont {P.~D.}\ \bibnamefont
  {Dernier}}\ and\ \bibinfo {author} {\bibfnamefont {M.}~\bibnamefont
  {Marezio}},\ }\bibfield  {title} {\bibinfo {title} {{Crystal Structure of the
  Low-Temperature Antiferromagnetic Phase of V$_2$O$_3$}},\ }\href
  {https://doi.org/10.1103/PhysRevB.2.3771} {\bibfield  {journal} {\bibinfo
  {journal} {Phys. Rev. B}\ }\textbf {\bibinfo {volume} {2}},\ \bibinfo {pages}
  {3771} (\bibinfo {year} {1970})}\BibitemShut {NoStop}%
\bibitem [{\citenamefont {Ueda}\ \emph {et~al.}(1980)\citenamefont {Ueda},
  \citenamefont {Kosuge},\ and\ \citenamefont {Kachi}}]{Ueda_1980}%
  \BibitemOpen
  \bibfield  {author} {\bibinfo {author} {\bibfnamefont {Y.}~\bibnamefont
  {Ueda}}, \bibinfo {author} {\bibfnamefont {K.}~\bibnamefont {Kosuge}},\ and\
  \bibinfo {author} {\bibfnamefont {S.}~\bibnamefont {Kachi}},\ }\bibfield
  {title} {\bibinfo {title} {{Phase diagram and some physical properties of
  V$_2$O$_{3+x}$ (0 $\leq$ $x$ $\leq$ 0.080)}},\ }\href
  {https://doi.org/https://doi.org/10.1016/0022-4596(80)90019-5} {\bibfield
  {journal} {\bibinfo  {journal} {Journal of Solid State Chemistry}\ }\textbf
  {\bibinfo {volume} {31}},\ \bibinfo {pages} {171} (\bibinfo {year}
  {1980})}\BibitemShut {NoStop}%
\bibitem [{\citenamefont {Singer}\ \emph {et~al.}(2018)\citenamefont {Singer},
  \citenamefont {Ramirez}, \citenamefont {Valmianski}, \citenamefont {Cela},
  \citenamefont {Hua}, \citenamefont {Kukreja}, \citenamefont {Wingert},
  \citenamefont {Kovalchuk}, \citenamefont {Glownia}, \citenamefont {Sikorski},
  \citenamefont {Chollet}, \citenamefont {Holt}, \citenamefont {Schuller},\
  and\ \citenamefont {Shpyrko}}]{Singer_2018}%
  \BibitemOpen
  \bibfield  {author} {\bibinfo {author} {\bibfnamefont {A.}~\bibnamefont
  {Singer}}, \bibinfo {author} {\bibfnamefont {J.~G.}\ \bibnamefont {Ramirez}},
  \bibinfo {author} {\bibfnamefont {I.}~\bibnamefont {Valmianski}}, \bibinfo
  {author} {\bibfnamefont {D.}~\bibnamefont {Cela}}, \bibinfo {author}
  {\bibfnamefont {N.}~\bibnamefont {Hua}}, \bibinfo {author} {\bibfnamefont
  {R.}~\bibnamefont {Kukreja}}, \bibinfo {author} {\bibfnamefont
  {J.}~\bibnamefont {Wingert}}, \bibinfo {author} {\bibfnamefont
  {O.}~\bibnamefont {Kovalchuk}}, \bibinfo {author} {\bibfnamefont {J.~M.}\
  \bibnamefont {Glownia}}, \bibinfo {author} {\bibfnamefont {M.}~\bibnamefont
  {Sikorski}}, \bibinfo {author} {\bibfnamefont {M.}~\bibnamefont {Chollet}},
  \bibinfo {author} {\bibfnamefont {M.}~\bibnamefont {Holt}}, \bibinfo {author}
  {\bibfnamefont {I.~K.}\ \bibnamefont {Schuller}},\ and\ \bibinfo {author}
  {\bibfnamefont {O.~G.}\ \bibnamefont {Shpyrko}},\ }\bibfield  {title}
  {\bibinfo {title} {{Nonequilibrium Phase Precursors during a Photoexcited
  Insulator-to-Metal Transition in V$_2$O$_3$}},\ }\href
  {https://doi.org/10.1103/PhysRevLett.120.207601} {\bibfield  {journal}
  {\bibinfo  {journal} {Phys. Rev. Lett.}\ }\textbf {\bibinfo {volume} {120}},\
  \bibinfo {pages} {207601} (\bibinfo {year} {2018})}\BibitemShut {NoStop}%
\bibitem [{\citenamefont {Ram\'{\i}rez}\ \emph {et~al.}(2016)\citenamefont
  {Ram\'{\i}rez}, \citenamefont {de~la Venta}, \citenamefont {Wang},
  \citenamefont {Saerbeck}, \citenamefont {Basaran}, \citenamefont {Batlle},\
  and\ \citenamefont {Schuller}}]{Ramirez_2016}%
  \BibitemOpen
  \bibfield  {author} {\bibinfo {author} {\bibfnamefont {J.~G.}\ \bibnamefont
  {Ram\'{\i}rez}}, \bibinfo {author} {\bibfnamefont {J.}~\bibnamefont {de~la
  Venta}}, \bibinfo {author} {\bibfnamefont {S.}~\bibnamefont {Wang}}, \bibinfo
  {author} {\bibfnamefont {T.}~\bibnamefont {Saerbeck}}, \bibinfo {author}
  {\bibfnamefont {A.~C.}\ \bibnamefont {Basaran}}, \bibinfo {author}
  {\bibfnamefont {X.}~\bibnamefont {Batlle}},\ and\ \bibinfo {author}
  {\bibfnamefont {I.~K.}\ \bibnamefont {Schuller}},\ }\bibfield  {title}
  {\bibinfo {title} {Collective mode splitting in hybrid heterostructures},\
  }\href {https://doi.org/10.1103/PhysRevB.93.214113} {\bibfield  {journal}
  {\bibinfo  {journal} {Phys. Rev. B}\ }\textbf {\bibinfo {volume} {93}},\
  \bibinfo {pages} {214113} (\bibinfo {year} {2016})}\BibitemShut {NoStop}%
\bibitem [{\citenamefont {Saerbeck}\ \emph {et~al.}(2014)\citenamefont
  {Saerbeck}, \citenamefont {de~la Venta}, \citenamefont {Wang}, \citenamefont
  {Ramírez}, \citenamefont {Erekhinsky}, \citenamefont {Valmianski},\ and\
  \citenamefont {Schuller}}]{Saerbeck_2014}%
  \BibitemOpen
  \bibfield  {author} {\bibinfo {author} {\bibfnamefont {T.}~\bibnamefont
  {Saerbeck}}, \bibinfo {author} {\bibfnamefont {J.}~\bibnamefont {de~la
  Venta}}, \bibinfo {author} {\bibfnamefont {S.}~\bibnamefont {Wang}}, \bibinfo
  {author} {\bibfnamefont {J.~G.}\ \bibnamefont {Ramírez}}, \bibinfo {author}
  {\bibfnamefont {M.}~\bibnamefont {Erekhinsky}}, \bibinfo {author}
  {\bibfnamefont {I.}~\bibnamefont {Valmianski}},\ and\ \bibinfo {author}
  {\bibfnamefont {I.~K.}\ \bibnamefont {Schuller}},\ }\bibfield  {title}
  {\bibinfo {title} {Coupling of magnetism and structural phase transitions by
  interfacial strain},\ }\href {https://doi.org/10.1557/jmr.2014.253}
  {\bibfield  {journal} {\bibinfo  {journal} {Journal of Materials Research}\
  }\textbf {\bibinfo {volume} {29}},\ \bibinfo {pages} {2353–2365} (\bibinfo
  {year} {2014})}\BibitemShut {NoStop}%
\bibitem [{\citenamefont {Valmianski}\ \emph {et~al.}(2021)\citenamefont
  {Valmianski}, \citenamefont {Rodríguez}, \citenamefont
  {Rodríguez-Álvarez}, \citenamefont {García~del Muro}, \citenamefont
  {Wolowiec}, \citenamefont {Kronast}, \citenamefont {Ramírez}, \citenamefont
  {Schuller}, \citenamefont {Labarta},\ and\ \citenamefont
  {Batlle}}]{Valmianski_2021}%
  \BibitemOpen
  \bibfield  {author} {\bibinfo {author} {\bibfnamefont {I.}~\bibnamefont
  {Valmianski}}, \bibinfo {author} {\bibfnamefont {A.~F.}\ \bibnamefont
  {Rodríguez}}, \bibinfo {author} {\bibfnamefont {J.}~\bibnamefont
  {Rodríguez-Álvarez}}, \bibinfo {author} {\bibfnamefont {M.}~\bibnamefont
  {García~del Muro}}, \bibinfo {author} {\bibfnamefont {C.}~\bibnamefont
  {Wolowiec}}, \bibinfo {author} {\bibfnamefont {F.}~\bibnamefont {Kronast}},
  \bibinfo {author} {\bibfnamefont {J.~G.}\ \bibnamefont {Ramírez}}, \bibinfo
  {author} {\bibfnamefont {I.~K.}\ \bibnamefont {Schuller}}, \bibinfo {author}
  {\bibfnamefont {A.}~\bibnamefont {Labarta}},\ and\ \bibinfo {author}
  {\bibfnamefont {X.}~\bibnamefont {Batlle}},\ }\bibfield  {title} {\bibinfo
  {title} {Driving magnetic domains at the nanoscale by interfacial
  strain-induced proximity},\ }\href {https://doi.org/10.1039/D0NR08253H}
  {\bibfield  {journal} {\bibinfo  {journal} {Nanoscale}\ }\textbf {\bibinfo
  {volume} {13}},\ \bibinfo {pages} {4985} (\bibinfo {year}
  {2021})}\BibitemShut {NoStop}%
\bibitem [{\citenamefont {{McGuire}}\ and\ \citenamefont
  {{Potter}}(1975)}]{McGuire_1975}%
  \BibitemOpen
  \bibfield  {author} {\bibinfo {author} {\bibfnamefont {T.}~\bibnamefont
  {{McGuire}}}\ and\ \bibinfo {author} {\bibfnamefont {R.}~\bibnamefont
  {{Potter}}},\ }\bibfield  {title} {\bibinfo {title} {Anisotropic
  magnetoresistance in ferromagnetic 3d alloys},\ }\href
  {https://doi.org/10.1109/TMAG.1975.1058782} {\bibfield  {journal} {\bibinfo
  {journal} {IEEE Transactions on Magnetics}\ }\textbf {\bibinfo {volume}
  {11}},\ \bibinfo {pages} {1018} (\bibinfo {year} {1975})}\BibitemShut
  {NoStop}%
\bibitem [{\citenamefont {Cullity}\ and\ \citenamefont
  {Graham}(2008{\natexlab{a}})}]{Cullity_2008}%
  \BibitemOpen
  \bibfield  {author} {\bibinfo {author} {\bibfnamefont {B.~D.}\ \bibnamefont
  {Cullity}}\ and\ \bibinfo {author} {\bibfnamefont {C.~D.}\ \bibnamefont
  {Graham}},\ }\bibinfo {title} {{Magnetic Anisotropy}},\ in\ \href
  {https://doi.org/https://doi.org/10.1002/9780470386323.ch7} {\emph {\bibinfo
  {booktitle} {Introduction to Magnetic Materials}}}\ (\bibinfo  {publisher}
  {John Wiley \& Sons, Ltd},\ \bibinfo {year} {2008})\ Chap.~\bibinfo {chapter}
  {7}, pp.\ \bibinfo {pages} {197 -- 239}\BibitemShut {NoStop}%
\bibitem [{\citenamefont {Coey}(2010{\natexlab{a}})}]{Coey_2010a}%
  \BibitemOpen
  \bibfield  {author} {\bibinfo {author} {\bibfnamefont {J.~M.~D.}\
  \bibnamefont {Coey}},\ }\bibinfo {title} {Ferromagnetism and exchange},\ in\
  \href {https://doi.org/10.1017/CBO9780511845000.006} {\emph {\bibinfo
  {booktitle} {Magnetism and Magnetic Materials}}}\ (\bibinfo  {publisher}
  {Cambridge University Press},\ \bibinfo {year} {2010})\ pp.\ \bibinfo {pages}
  {128 -- 194}\BibitemShut {NoStop}%
\bibitem [{\citenamefont {Coey}(2010{\natexlab{b}})}]{Coey_2010b}%
  \BibitemOpen
  \bibfield  {author} {\bibinfo {author} {\bibfnamefont {J.~M.~D.}\
  \bibnamefont {Coey}},\ }\bibinfo {title} {Micromagnetism, domains and
  hysteresis},\ in\ \href {https://doi.org/10.1017/CBO9780511845000.008} {\emph
  {\bibinfo {booktitle} {Magnetism and Magnetic Materials}}}\ (\bibinfo
  {publisher} {Cambridge University Press},\ \bibinfo {year} {2010})\ pp.\
  \bibinfo {pages} {231 -- 263}\BibitemShut {NoStop}%
\bibitem [{\citenamefont {Urban}\ \emph {et~al.}(2001)\citenamefont {Urban},
  \citenamefont {Woltersdorf},\ and\ \citenamefont {Heinrich}}]{Urban_2001}%
  \BibitemOpen
  \bibfield  {author} {\bibinfo {author} {\bibfnamefont {R.}~\bibnamefont
  {Urban}}, \bibinfo {author} {\bibfnamefont {G.}~\bibnamefont {Woltersdorf}},\
  and\ \bibinfo {author} {\bibfnamefont {B.}~\bibnamefont {Heinrich}},\
  }\bibfield  {title} {\bibinfo {title} {{Gilbert Damping in Single and
  Multilayer Ultrathin Films: Role of Interfaces in Nonlocal Spin Dynamics}},\
  }\href {https://doi.org/10.1103/PhysRevLett.87.217204} {\bibfield  {journal}
  {\bibinfo  {journal} {Phys. Rev. Lett.}\ }\textbf {\bibinfo {volume} {87}},\
  \bibinfo {pages} {217204} (\bibinfo {year} {2001})}\BibitemShut {NoStop}%
\bibitem [{\citenamefont {Tannous}\ and\ \citenamefont
  {Gieraltowski}(2008)}]{Tannous_2008}%
  \BibitemOpen
  \bibfield  {author} {\bibinfo {author} {\bibfnamefont {C.}~\bibnamefont
  {Tannous}}\ and\ \bibinfo {author} {\bibfnamefont {J.}~\bibnamefont
  {Gieraltowski}},\ }\bibfield  {title} {\bibinfo {title} {{The
  Stoner{\textendash}Wohlfarth model of ferromagnetism}},\ }\href
  {https://doi.org/10.1088/0143-0807/29/3/008} {\bibfield  {journal} {\bibinfo
  {journal} {European Journal of Physics}\ }\textbf {\bibinfo {volume} {29}},\
  \bibinfo {pages} {475} (\bibinfo {year} {2008})}\BibitemShut {NoStop}%
\bibitem [{\citenamefont {Cullity}\ and\ \citenamefont
  {Graham}(2008{\natexlab{b}})}]{Cullity_2008b}%
  \BibitemOpen
  \bibfield  {author} {\bibinfo {author} {\bibfnamefont {B.~D.}\ \bibnamefont
  {Cullity}}\ and\ \bibinfo {author} {\bibfnamefont {C.~D.}\ \bibnamefont
  {Graham}},\ }\bibinfo {title} {{Magnetostriction and the Effects of
  Stress}},\ in\ \href
  {https://doi.org/https://doi.org/10.1002/9780470386323.ch8} {\emph {\bibinfo
  {booktitle} {Introduction to Magnetic Materials}}}\ (\bibinfo  {publisher}
  {John Wiley \& Sons, Ltd},\ \bibinfo {year} {2008})\ Chap.~\bibinfo {chapter}
  {8}, pp.\ \bibinfo {pages} {241--273}\BibitemShut {NoStop}%
\bibitem [{\citenamefont {Asai}\ \emph {et~al.}(2016)\citenamefont {Asai},
  \citenamefont {Ota}, \citenamefont {Namazu}, \citenamefont {Takenobu},
  \citenamefont {Koyama},\ and\ \citenamefont {Chiba}}]{Asai_2016}%
  \BibitemOpen
  \bibfield  {author} {\bibinfo {author} {\bibfnamefont {R.}~\bibnamefont
  {Asai}}, \bibinfo {author} {\bibfnamefont {S.}~\bibnamefont {Ota}}, \bibinfo
  {author} {\bibfnamefont {T.}~\bibnamefont {Namazu}}, \bibinfo {author}
  {\bibfnamefont {T.}~\bibnamefont {Takenobu}}, \bibinfo {author}
  {\bibfnamefont {T.}~\bibnamefont {Koyama}},\ and\ \bibinfo {author}
  {\bibfnamefont {D.}~\bibnamefont {Chiba}},\ }\bibfield  {title} {\bibinfo
  {title} {{Stress-induced large anisotropy field modulation in Ni films
  deposited on a flexible substrate}},\ }\href
  {https://doi.org/10.1063/1.4961564} {\bibfield  {journal} {\bibinfo
  {journal} {Journal of Applied Physics}\ }\textbf {\bibinfo {volume} {120}},\
  \bibinfo {pages} {083906} (\bibinfo {year} {2016})}\BibitemShut {NoStop}%
\bibitem [{\citenamefont {Bozorth}(1946)}]{Bozorth_1946}%
  \BibitemOpen
  \bibfield  {author} {\bibinfo {author} {\bibfnamefont {R.~M.}\ \bibnamefont
  {Bozorth}},\ }\bibfield  {title} {\bibinfo {title} {{Magnetoresistance and
  Domain Theory of Iron-Nickel Alloys}},\ }\href
  {https://doi.org/10.1103/PhysRev.70.923} {\bibfield  {journal} {\bibinfo
  {journal} {Phys. Rev.}\ }\textbf {\bibinfo {volume} {70}},\ \bibinfo {pages}
  {923} (\bibinfo {year} {1946})}\BibitemShut {NoStop}%
\bibitem [{\citenamefont {Smit}(1951)}]{Smit_1951}%
  \BibitemOpen
  \bibfield  {author} {\bibinfo {author} {\bibfnamefont {J.}~\bibnamefont
  {Smit}},\ }\bibfield  {title} {\bibinfo {title} {Magnetoresistance of
  ferromagnetic metals and alloys at low temperatures},\ }\href
  {https://doi.org/https://doi.org/10.1016/0031-8914(51)90117-6} {\bibfield
  {journal} {\bibinfo  {journal} {Physica}\ }\textbf {\bibinfo {volume} {17}},\
  \bibinfo {pages} {612} (\bibinfo {year} {1951})}\BibitemShut {NoStop}%
\bibitem [{\citenamefont {McLeod}\ \emph {et~al.}(2017)\citenamefont {McLeod},
  \citenamefont {van Heumen}, \citenamefont {Ramirez}, \citenamefont {Wang},
  \citenamefont {Saerbeck}, \citenamefont {Guenon}, \citenamefont {Goldflam},
  \citenamefont {Anderegg}, \citenamefont {Kelly}, \citenamefont {Mueller},
  \citenamefont {Liu}, \citenamefont {Schuller},\ and\ \citenamefont
  {Basov}}]{McLeod_2017}%
  \BibitemOpen
  \bibfield  {author} {\bibinfo {author} {\bibfnamefont {A.~S.}\ \bibnamefont
  {McLeod}}, \bibinfo {author} {\bibfnamefont {E.}~\bibnamefont {van Heumen}},
  \bibinfo {author} {\bibfnamefont {J.~G.}\ \bibnamefont {Ramirez}}, \bibinfo
  {author} {\bibfnamefont {S.}~\bibnamefont {Wang}}, \bibinfo {author}
  {\bibfnamefont {T.}~\bibnamefont {Saerbeck}}, \bibinfo {author}
  {\bibfnamefont {S.}~\bibnamefont {Guenon}}, \bibinfo {author} {\bibfnamefont
  {M.}~\bibnamefont {Goldflam}}, \bibinfo {author} {\bibfnamefont
  {L.}~\bibnamefont {Anderegg}}, \bibinfo {author} {\bibfnamefont
  {P.}~\bibnamefont {Kelly}}, \bibinfo {author} {\bibfnamefont
  {A.}~\bibnamefont {Mueller}}, \bibinfo {author} {\bibfnamefont {M.~K.}\
  \bibnamefont {Liu}}, \bibinfo {author} {\bibfnamefont {I.~K.}\ \bibnamefont
  {Schuller}},\ and\ \bibinfo {author} {\bibfnamefont {D.~N.}\ \bibnamefont
  {Basov}},\ }\bibfield  {title} {\bibinfo {title} {{Nanotextured phase
  coexistence in the correlated insulator V$_2$O$_3$}},\ }\href@noop {}
  {\bibfield  {journal} {\bibinfo  {journal} {Nat. Phys.}\ }\textbf {\bibinfo
  {volume} {13}},\ \bibinfo {pages} {80} (\bibinfo {year} {2017})}\BibitemShut
  {NoStop}%
\bibitem [{\citenamefont {Davies}\ \emph {et~al.}(2004)\citenamefont {Davies},
  \citenamefont {Hellwig}, \citenamefont {Fullerton}, \citenamefont {Denbeaux},
  \citenamefont {Kortright},\ and\ \citenamefont {Liu}}]{Davies_2004}%
  \BibitemOpen
  \bibfield  {author} {\bibinfo {author} {\bibfnamefont {J.~E.}\ \bibnamefont
  {Davies}}, \bibinfo {author} {\bibfnamefont {O.}~\bibnamefont {Hellwig}},
  \bibinfo {author} {\bibfnamefont {E.~E.}\ \bibnamefont {Fullerton}}, \bibinfo
  {author} {\bibfnamefont {G.}~\bibnamefont {Denbeaux}}, \bibinfo {author}
  {\bibfnamefont {J.~B.}\ \bibnamefont {Kortright}},\ and\ \bibinfo {author}
  {\bibfnamefont {K.}~\bibnamefont {Liu}},\ }\bibfield  {title} {\bibinfo
  {title} {Magnetization reversal of copt multilayers: Microscopic origin of
  high-field magnetic irreversibility},\ }\href
  {https://doi.org/10.1103/PhysRevB.70.224434} {\bibfield  {journal} {\bibinfo
  {journal} {Phys. Rev. B}\ }\textbf {\bibinfo {volume} {70}},\ \bibinfo
  {pages} {224434} (\bibinfo {year} {2004})}\BibitemShut {NoStop}%
\bibitem [{\citenamefont {Abadias}\ \emph {et~al.}(2018)\citenamefont
  {Abadias}, \citenamefont {Chason}, \citenamefont {Keckes}, \citenamefont
  {Sebastiani}, \citenamefont {Thompson}, \citenamefont {Barthel},
  \citenamefont {Doll}, \citenamefont {Murray}, \citenamefont {Stoessel},\ and\
  \citenamefont {Martinu}}]{Abadias_2018}%
  \BibitemOpen
  \bibfield  {author} {\bibinfo {author} {\bibfnamefont {G.}~\bibnamefont
  {Abadias}}, \bibinfo {author} {\bibfnamefont {E.}~\bibnamefont {Chason}},
  \bibinfo {author} {\bibfnamefont {J.}~\bibnamefont {Keckes}}, \bibinfo
  {author} {\bibfnamefont {M.}~\bibnamefont {Sebastiani}}, \bibinfo {author}
  {\bibfnamefont {G.~B.}\ \bibnamefont {Thompson}}, \bibinfo {author}
  {\bibfnamefont {E.}~\bibnamefont {Barthel}}, \bibinfo {author} {\bibfnamefont
  {G.~L.}\ \bibnamefont {Doll}}, \bibinfo {author} {\bibfnamefont {C.~E.}\
  \bibnamefont {Murray}}, \bibinfo {author} {\bibfnamefont {C.~H.}\
  \bibnamefont {Stoessel}},\ and\ \bibinfo {author} {\bibfnamefont
  {L.}~\bibnamefont {Martinu}},\ }\bibfield  {title} {\bibinfo {title} {{Review
  Article: Stress in thin films and coatings: Current status, challenges, and
  prospects}},\ }\href {https://doi.org/10.1116/1.5011790} {\bibfield
  {journal} {\bibinfo  {journal} {Journal of Vacuum Science \& Technology A}\
  }\textbf {\bibinfo {volume} {36}},\ \bibinfo {pages} {020801} (\bibinfo
  {year} {2018})}\BibitemShut {NoStop}%
\bibitem [{\citenamefont {Institute~of
  Materials~Engineering}(1997)}]{IMMA_1997}%
  \BibitemOpen
  \bibfield  {author} {\bibinfo {author} {\bibfnamefont {A.}~\bibnamefont
  {Institute~of Materials~Engineering}},\ }\href@noop {} {\emph {\bibinfo
  {title} {IMMA handbook of engineering materials}}},\ \bibinfo {edition}
  {5th}\ ed.\ (\bibinfo  {publisher} {"Parkville, Vic.: Institute of Metals and
  Materials Australasia"},\ \bibinfo {address} {Parkville, Victoria},\ \bibinfo
  {year} {1997})\BibitemShut {NoStop}%
\end{thebibliography}%


\providecommand{\noopsort}[1]{}\providecommand{\singleletter}[1]{#1}%
%
\appendix*
\section{Supplemental Material}
\setcounter{figure}{0}
\renewcommand{\figurename}{Fig.}
\renewcommand{\thefigure}{S\arabic{figure}}
\begin{figure*}[t]
   \includegraphics[scale = 1, trim= 0cm 0cm 0cm 0cm, clip=true]{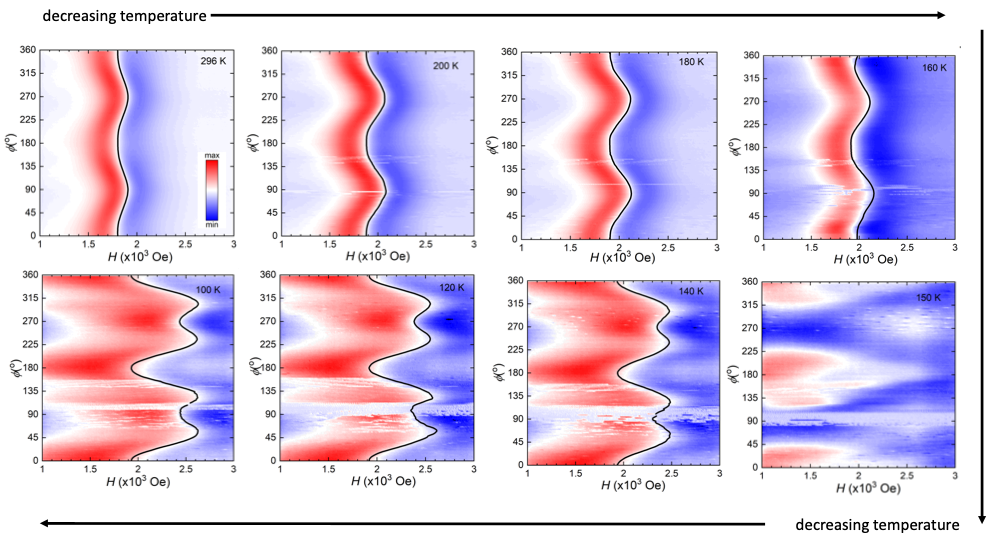}
\caption{\label{figS1} (Color online) In-plane angular-dependent ferromagnetic resonance (FMR) measurements of the V$_2$O$_3$/Ni bilayer sample $pS1$ with a polycrystalline Ni layer. The angular dependence of the resonance field $H_R$ (solid black curve) was determined from fits of a dynamical model for the intensity to the FMR data at each angle $\phi$. The dynamical model for the intensity $I(H )$  is given by the derivative of a Lorentzian: $I(H )= -C\cdot(H - H_R)/(4\cdot(H - H_R)^2 + {\Delta}^2)^2 + B$, where $H_R$ is the resonance field, $\Delta$ is the line width, $C$ is a scaling factor, and $B$ is a background intensity. The disorder in the magnetic structure at 150 K prevented reasonable fits of the model to the FMR data. The blue to red contrast indicates the low to high signal intensity as a function of magnetic field applied parallel to the Ni film surface. The angle between the applied magnetic field and the in-plane magnetic easy axis is represented by $\phi$ along the $y$-axis. The data shown was taken upon cooling from 296 to 100 K.} 
\end{figure*}
\begin{figure*}[t]
   \includegraphics[scale = 1, trim= 0cm 0cm 0cm 0cm, clip=true]{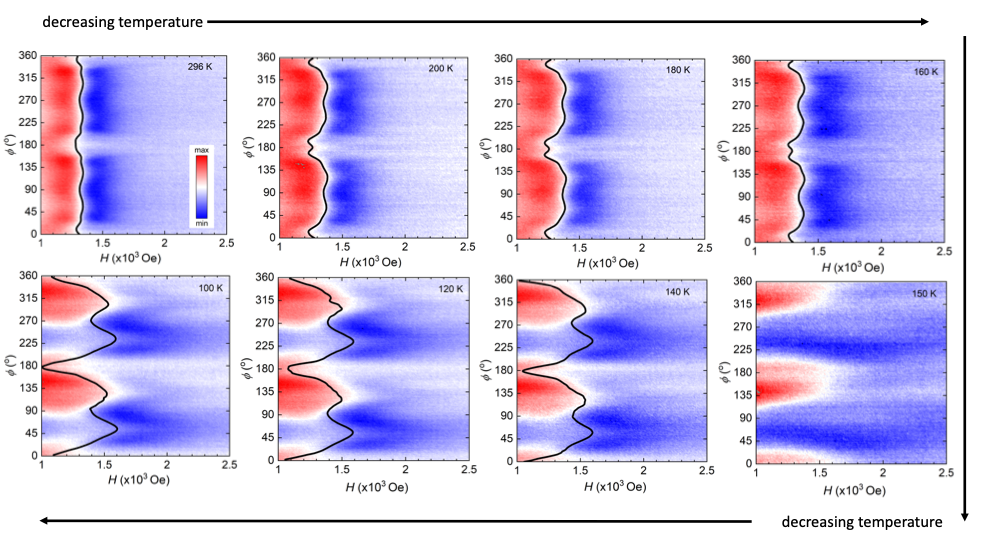}
\caption{\label{figS2} (Color online) In-plane angular-dependent ferromagnetic resonance (FMR) measurements of V$_2$O$_3$/Ni bilayer sample $tS2$ at low microwave power of 1 mW. The high temperature (HT) magnetic anisotropy vanishes at or above a cavity temperature of 150 K. The angular dependence of the resonance field $H_R$ (solid black curve) was determined from fits of a dynamical model for the intensity to the FMR data at each angle $\phi$. The dynamical model for the intensity $I(H )$ is given by the derivative of a Lorentzian: $I(H )= -C\cdot(H - H_R)/(4\cdot(H - H_R)^2 + {\Delta}^2)^2 + B$, where $H_R$ is the resonance field, $\Delta$ is the line width, $C$ is a scaling factor, and $B$ is a background intensity. The disorder in the magnetic structure at 150 K prevented reasonable fits of the model to the FMR data. The blue to red contrast indicates the low to high signal intensity as a function of magnetic field applied parallel to the Ni film surface. The angle between the applied magnetic field and the in-plane magnetic easy axis is represented by $\phi$ along the $y$-axis. The data shown was taken upon cooling from 296 to 100 K.} 
\end{figure*}
\begin{figure*}[t]
   \includegraphics[scale = 1, trim= 0cm 0cm 0cm 0cm, clip=true]{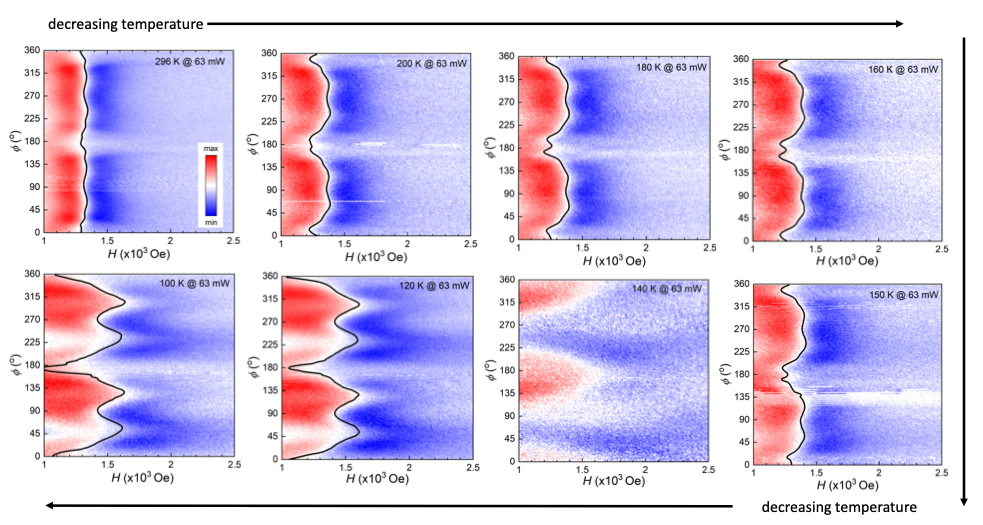}
\caption{\label{figS3} (Color online) In-plane angular-dependent ferromagnetic resonance (FMR) measurements of the V$_2$O$_3$/Ni bilayer sample $tS2$ at high microwave power of 63 mW. Increasing the microwave power has little to no effect on the resonance field $H_R$ or line width $\Delta$ as compared to the FMR measurements at a microwave power of 1 mW (see Fig.~\ref{figS2}). However, significant heating of the sample (of approximately 10 K) occurs at this power of 63 mW where the high temperature (HT) magnetic anisotropy of Ni in proximity to V$_2$O$_3$ vanishes at or above 140 K and reemerges at low temperature (LT) in the V$_2$O$_3$ monoclinic phase. At 1 mW, the high temperature (HT) magnetic anisotropy vanishes at or above a cavity temperature of 150 K (see Fig.~\ref{figS2}). The 10 K discrepancy in the V$_2$O$_3$ SPT temperature for the FMR measurements at 1 mW and 63 mW suggests the sample is being heated considerably at higher power. The angular dependence of the resonance field H$_R$ (solid black curve) was determined from fits of a dynamical model for the intensity to the FMR data at each angle $\phi$. The dynamical model for the intensity $I(H )$ is given by the derivative of a Lorentzian: $I(H )= -C\cdot(H - H_R)/(4\cdot(H - H_R)^2 + {\Delta}^2)^2 + B$, where $H_R$ is the resonance field, $\Delta$ is the line width, $C$ is a scaling factor, and $B$ is a background intensity. The disorder in the magnetic structure at 150 K prevented reasonable fits of the model to the FMR data. The blue to red contrast indicates the low to high signal intensity as a function of magnetic field applied parallel to the Ni film surface. The angle between the applied magnetic field and the in-plane magnetic easy axis is represented by $\phi$ along the $y$-axis. The data shown was taken upon cooling from 296 to 100 K.} 
\end{figure*}

\begin{figure*}[t]
   \includegraphics[scale = 1, trim= 0cm 0cm 0cm 0cm, clip=true]{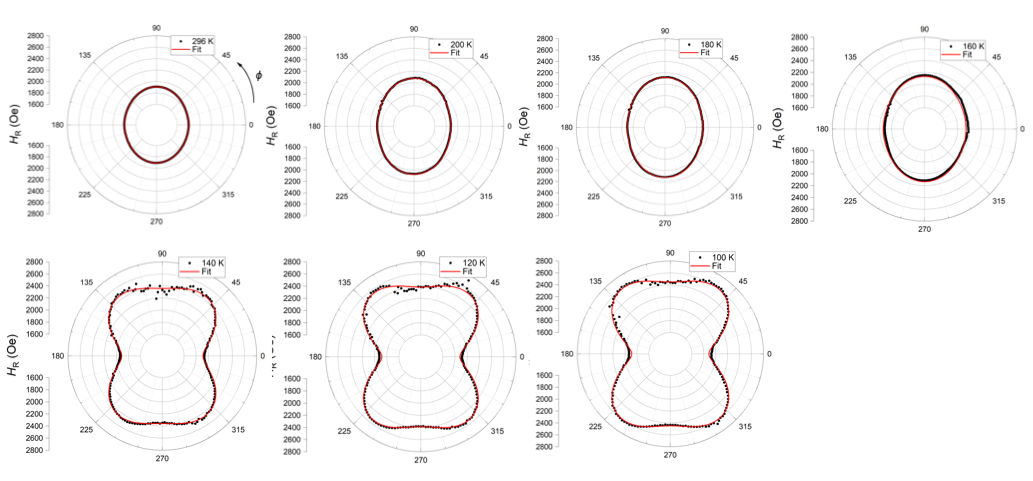}
\caption{\label{figS4}Polar-plot representations of the resonance field lines $H_R(\phi)$ that were determined from the FMR measurements for sample $pS1$ at temperatures from 296 K down to 100 K. The red curves are fits of the anisotropy energy $E_a$ (see Equation 1 in main text) to the resonance field lines $H_R(\phi)$ (black circles).}
\end{figure*} 

\begin{figure*}[t]
   \includegraphics[scale = 1, trim= 0cm 0cm 0cm 0cm, clip=true]{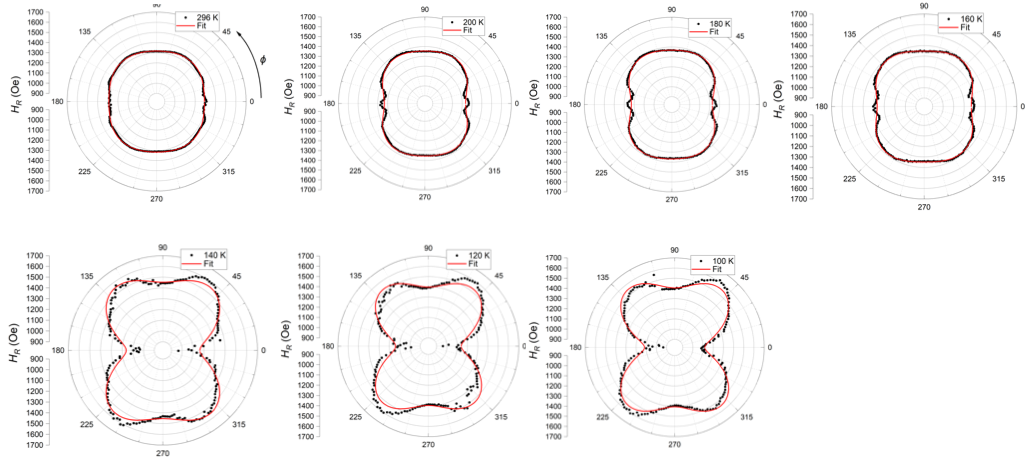}
\caption{\label{figS5}Polar-plot representations of the resonance field lines $H_R(\phi)$ that were determined from the FMR measurements for sample $tS2$ at temperatures from 296 K down to 100 K. The red curves are fits of the anisotropy energy $E_a$ (see Equation 1 in main text) to the resonance field lines $H_R(\phi)$ (black circles).}
\end{figure*} 

\begin{figure*}[t]
   \includegraphics[scale = 1, trim= 0cm 0cm 0cm 0cm, clip=true]{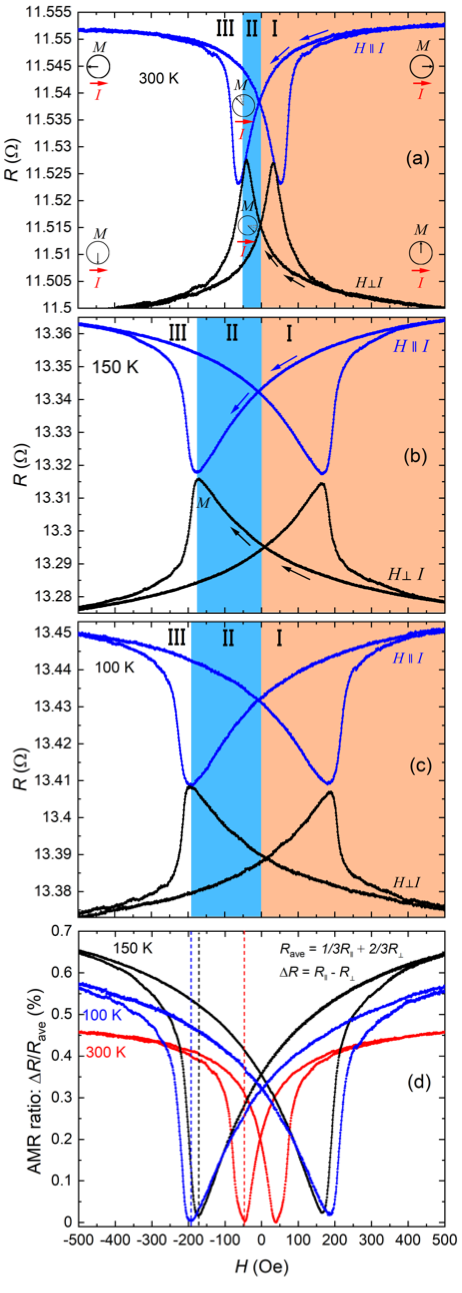}
\caption{\label{figS6}The longitudinal and transverse magnetoresistance (MR) for the $tS2$ sample at (a) 300 K,  (b) 150 K, and (c) 100 K. The blue (black) curves show the MR when the applied magnetic field is parallel $H_{\parallel}$ (or perpendicular $H_{\perp}$) to the direction of the electric current $I$. (d) The anisotropic magnetoresistance (AMR) ratio $\Delta$$R$/$R_{ave}$ with $\Delta$$R$ $=$ $R_{\parallel}$ $-$ $R_{\perp}$ and $R_{ave}$ $=$ 1/3$R_{\parallel}$ + 2/3$R_{\perp}$. Note that the normalized AMR effect is largest at 150 K near the temperature of the SPT in V$_2$O$_3$.}
\end{figure*}

\begin{figure*}[t]
   \includegraphics[scale = 1, trim= 0cm 0cm 0cm 0cm, clip=true]{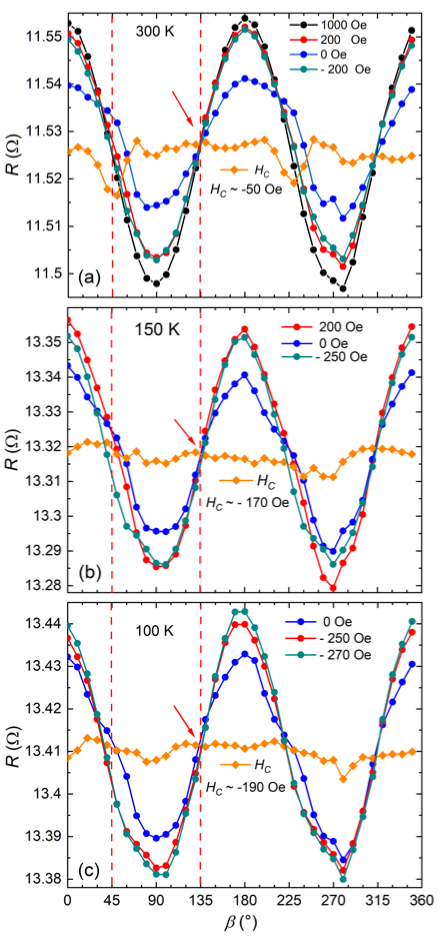}
\caption{\label{figS7} The angular dependence of the magnetoresistance (MR) at various field for the $tS2$ sample at temperatures (a) 300 K, (b) 150 K, and (c) 100 K above and below the SPT in V$_2$O$_3$. The red arrows and vertical dashed lines highlight the intersection of $R$ at the coercive field $H_c$ with $R$ in a sufficiently strong applied magnetic field directed at 45\degree, 135\degree, 225\degree, 315\degree.}
\end{figure*}

\end{document}